\documentclass[%
reprint,
superscriptaddress,
amsmath, amssymb,
aps,
prb,
]{revtex4-2}

\usepackage[utf8]{inputenc}
\usepackage{standalone}
\usepackage{graphicx}
\usepackage{dcolumn}
\usepackage{multirow}
\usepackage{diagbox}
\usepackage{bm}
\usepackage{braket}
\usepackage{dsfont}
\usepackage{color} 

\usepackage[normalem]{ulem}
\usepackage{svg}
\usepackage[%
  colorlinks=true,
  urlcolor=blue,
  linkcolor=blue,
  citecolor=blue
]{hyperref}

\begin{document}

\title{Surface-state engineering for nonlinear charge and spin photocurrent generation}
\date{\today}
\author{Javier Sivianes}
\affiliation{Centro de F\'{i}sica de Materiales (CSIC-UPV/EHU), 20018, Donostia-San Sebasti\'{a}n, Spain}

\author{Peio Garcia-Goiricelaya}
\affiliation{Universidad del País Vasco/Euskal Herriko Unibertsitatea (UPV/EHU)}
\author{Daniel Hernangómez-Pérez}
\affiliation{CIC nanoGUNE BRTA, Tolosa Hiribidea 76, 20018 San Sebasti\'an, Spain}
\author{Julen Iba\~{n}ez-Azpiroz}%
\affiliation{Centro de F\'{i}sica de Materiales (CSIC-UPV/EHU), 20018, Donostia-San Sebasti\'{a}n, Spain}
\affiliation{IKERBASQUE, Basque Foundation for Science, 48009 Bilbao, Spain}
\affiliation{Donostia International Physics Center (DIPC), 20018
Donostia-San Sebasti\'{a}n, Spain}

\begin{abstract}
We systematically explore a pathway for generating nonlinear charge and spin photocurrents using spin-orbit-split surface states.
This mechanism enables net charge and spin flow along the surface plane even in centrosymmetric bulk environments like the Rashba prototype Au(111), where we establish the key principles 
by combining model predictions with density functional calculations.
We further identify the Tl/Si(111) surface, characterized by strong non-Rashba spin-orbit coupling, as a prime candidate for experimental validation; with slight doping, it develops metallic spin-orbit-split states featuring  remarkable relativistic properties  
while the bulk remains semiconducting. Our non-linear simulations reveal distinct angular signatures and magnitudes comparable to bulk ferroelectrics, highlighting the potential of surface-state photocurrents for low-bias optoelectronic applications. Moreover, the intricate spin polarization of surface states opens new possibilities 
as a nonlinear spin filter,
providing a far more versatile platform than the spin Hall effect.
\end{abstract}

\maketitle

\begin{figure}
    \centering
    \includegraphics[width=0.3\textwidth]{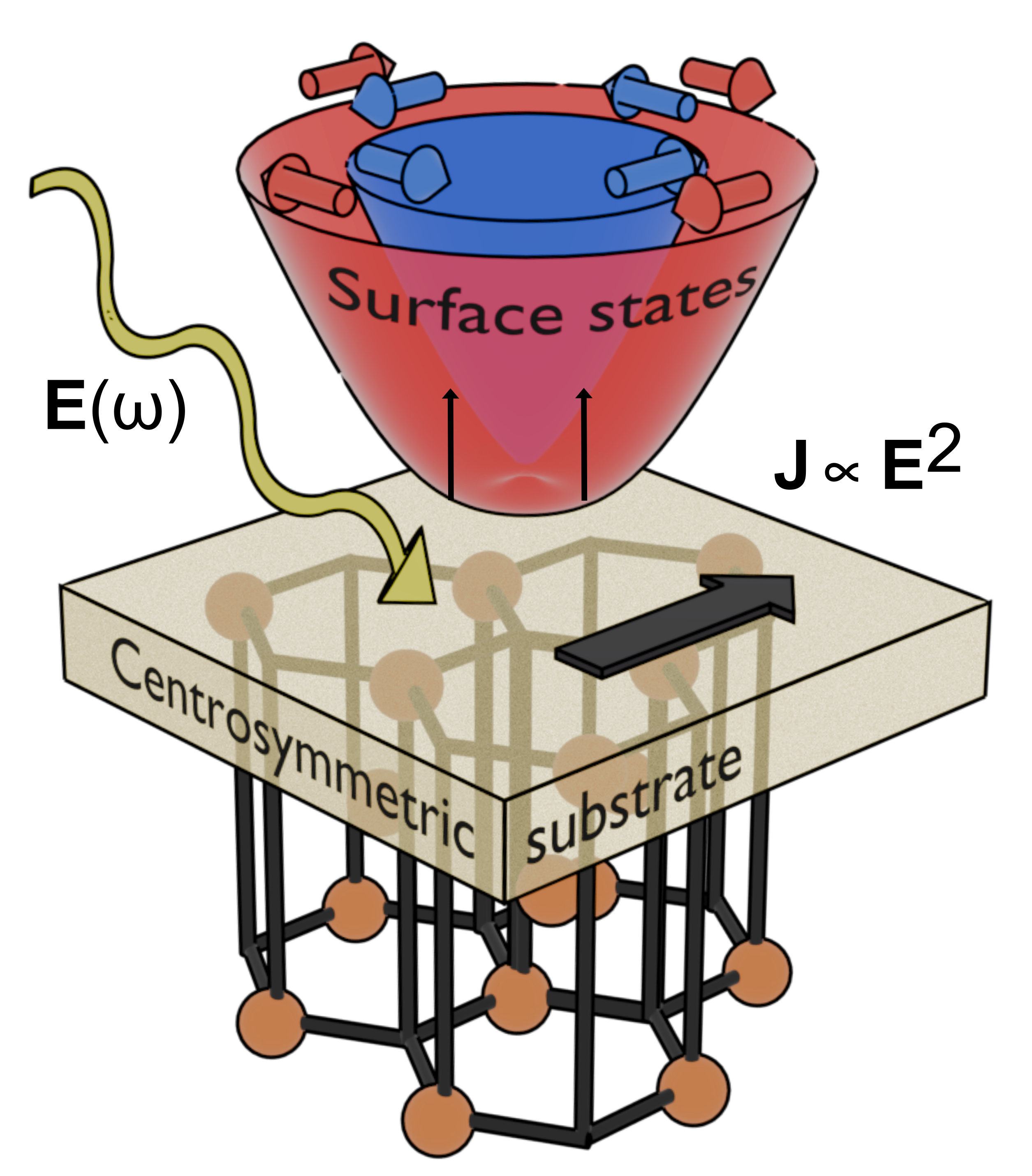}
    \caption{Schematic representation of the nonlinear phototransport concept; incoming light generates current through a quadratic optical response in the surface of a centrosymmetric crystal. The characteristic energy dispersion and spin texture of the Rashba surface states are displayed. The underlying represented surface and crystal structure corresponds to Au(111).}
    \label{fig:intro}
\end{figure}

Materials that lack a center of inversion in their crystalline structure exhibit a  distinctive optical effect known as the bulk photovoltaic effect (BVPE)~\cite {belinicher-jetp82,sturman-book92,baltz-prb81}. 
This phenomenon is described by a quadratic optical response that generates a net DC photocurrent in homogeneous materials. 
The effect  has enormous potential for applications in interface-free solar cell technologies, which  represents a notable advantage over the standard photovoltaics  based on pn junctions.
In recent years, materials such as Weyl semimetals~\cite{Ma2019,osterhoudt2019colossal} or transition-metal dichalcogenide nanotubes~\cite{zhang_enhanced_2019,PhysRevB.108.165418} have exhibited a striking enhancement of the BVPE, approaching  competitive figures of merit.
In addition, recent works suggest that the spin-dependent counterpart of the effect \cite{PhysRevB.105.045201} might pave the way for novel applications, \textit{e.g.}  
as pure spin current generators~\cite{xu_pure_2021}
or as a knob of unconventional magnetic phases~\cite{dong2024crystalsymmetryselectedpure}.

A significant challenge in identifying materials 
with an enhanced BVPE is the necessity of broken inversion symmetry, 
which greatly limits the pool of viable candidates. 
However, even in bulk centrosymmetric materials inversion symmetry is inherently broken at their \emph{surface}, enabling symmetry-allowed quadratic optical responses~\cite{PhysRevB.90.205310, PhysRevB.84.195326}.
In this scenario, the nonlinear response is expected to be most prominent in states highly localized around the surface, particularly in electronic transitions involving \emph{surface states}. Given the strong ties between surface states and relativistic effects such as the spin-orbit interaction, this approach provides an excellent platform to explore the influence of spin-orbit physics on nonlinear optical absorption. 

In this study, we provide a quantitative investigation of nonlinear photocurrents generated  by spin-orbit split surface states. 
As illustrated in Fig.~\ref{fig:intro}, this mechanism enables a net DC charge and spin flow along the surface plane, even within a bulk environment that maintains centrosymmetry. 
Despite the appeal of such mechanism, 
previous theoretical studies have been
limited to the specific context of 
surface states in topological insulators 
and explored solely at the model  level~\cite{PhysRevB.95.035134,PhysRevB.107.L161403}.
Here, we substantiate our findings through density functional theory (DFT) calculations, 
offering realistic estimates for spin and charge 
photocurrent magnitudes in specific materials. 
First, we establish the fundamental principles in
the Au(111) surface featuring the prototype Rashba spin-orbit split states as sketched in Fig.~\ref{fig:intro}. 
Next, we identify Tl/Si(111) as an ideal candidate for experimental realization, providing clear and 
measurable signatures of surface-state charge photocurrents. Finally, we demonstrate that its unique surface-state spin-polarization structure near the K and K$^{\prime}$ valleys provides the system with the functionality of a nonlinear spin filter.

We begin by describing the quadratic optical contributions to the BVPE. 
Under a monochromatic light field $\bm{E}(t)=\bm{E}(\omega)e^{i\omega t}+ \bm{E}(-\omega)e^{-i\omega t}$, the generated 
DC charge photocurrent ($\omega = 0$)
has two sources, namely the shift and injection currents.
In a time-reversal invariant material, it can be written as
\begin{equation}\label{eq:definition}
    J^{a}(0) = \sigma^{abc}\text{Re}[E^{b}(\omega)E^{c}(-\omega)] + \tau\cdot\eta^{abc} \text{Im}[E^{b}(\omega)E^{c}(-\omega)]
\end{equation}
with $a$, $b$ and $c$ Cartesian indexes.
In Eq. \eqref{eq:definition}, $\sigma^{abc}$ and $\eta^{abc}$ stand for the shift and injection 
photoconductivities, respectively. The former 
describes an intrinsic contribution to the current
originating from real-space shifts undergone by electrons upon photoexcitation,
which is symmetric in the electric field indexes.
In contrast, $\eta^{abc}$ is anti-symmetric in  $bc$, 
and its generation rate is proportional to the electronic relaxation time $\tau$.
These contributions were intensively studied a 
few decades ago~\cite {belinicher-jetp82,sturman-book92,baltz-prb81}, and 
their modern length-gauge expressions employed in this work 
were derived in Ref.~\cite{PhysRevB.61.5337}
(see more details in the Supporting Information \cite{sup}).

\begin{figure*}[!t]
\includegraphics[width=\textwidth]{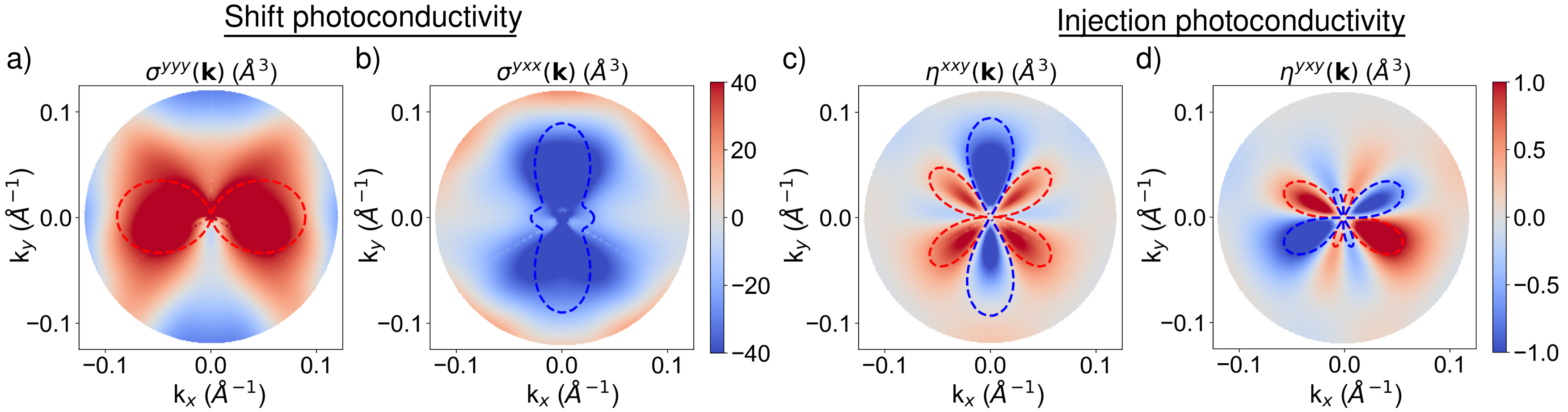}
\caption{Components of \textbf{k}-resolved transition matrix elements of the shift [(a) and (b)] and injection [(c) and (d)] photoconductivities, showing with a colormap the calculated values for the Au(111) surface states. The dashed lines correspond to the analytical results from the Rashba model [Eqs. \ref{eq:I} and \ref{eq:eta} correspond to (b) and (c)].}\label{fig:kslice}
\end{figure*}

As a specific material realization, 
we first consider Au(111), a system where Lashell et al. famously reported evidence of spin-split surface states driven by spin-orbit coupling~\cite{PhysRevLett.77.3419}.
Since then, it has become a paradigmatic example of the Rashba two-dimensional electron gas~\cite{sobihlmayer,au111-exp,PhysRevB.88.125404}. 
Fig.~\ref{fig:intro} pictorially depicts the band structure near the $\Gamma$ point, highlighting its characteristic spin-orbit splitting and the purely in-plane spin texture.
Unlike bulk Au, Au(111) belongs to the noncentrosymmetric point group 3m, enabling symmetry-allowed quadratic optical responses at the surface.
Table \ref{tab:sym} displays the permitted (independent) components  
which can generate current under linearly polarized light (LPL) and 
circularly polarized light (CPL) by the quadratic shift and injection 
mechanisms.

\begin{table}
\centering
\caption{Independent components of the charge and spin shift and injection photoconductivity tensors for the 
3$m$ point group describing Au(111) and Tl/Si(111) surfaces.\vspace{0.1cm}}
\begin{tabular}{|c|c|c|}
\hline
     & LPL & CPL \\ \hline
     Charge shift & ${yyy}$ & $-$ \\ \hline
     Charge injection & $-$ & $-$ \\ \hline
     Spin shift & $-$ & ${x;xxy}$, ${x;xxz}$, ${z;xyz}$ \\ \hline
     \multirow{2}{*}{Spin injection} & ${x;xxy}$, ${x;xxz}$, ${x;yxx}$, & \multirow{2}{*}{$-$} \\ 
      &  ${x;yzz}$, ${z;xxx}$, ${z;xyz}$ & \\ \hline
\end{tabular}
\label{tab:sym}
\end{table}

We focus on excitations that connect the spin-orbit split surface states of Au(111) (see 
Fig.~\ref{fig:intro}) and employ the Rashba model to gain  insight into the underlying physical properties. 
%
At linear order in the crystal momentum $\mathbf{k}$, the Rashba model is isotropic.
However, it can be extended with higher-order terms that comply with specific point group symmetries. In this work, we incorporate terms up to third order in $\mathbf{k}$ that transform according to the noncentrosymmetric  point group $3m$ of Au(111), as originally proposed in Ref. \cite{PhysRevLett.103.266801}.
The $k \cdot p$ Hamiltonian in polar coordinates ($\rho,\phi$) reads 
\begin{align}
    H &= \frac{\rho^{2}}{2m}\mathcal{I} - (\alpha + \gamma \rho^{2}) \rho \sin(\phi) \sigma_x  \nonumber \\  
     & + (\alpha + \gamma \rho^{2}) \rho \cos(\phi) \sigma_y+ \beta \rho^{3} \cos(3\phi) \sigma_z.
\end{align}
It is composed by the isotropic linear term characterized by the Rashba coupling constant $\alpha$, a third order isotropic correction with coefficient $\gamma$, and the hexagonal warping term characterized by $\beta$ \cite{PhysRevLett.103.266801}.
%
The inclusion of hexagonal warping is essential to break the symmetry of the otherwise isotropic low energy Hamiltonian, allowing in-plane surface currents.
Since for Au(111) the Rashba model is only valid in a small region close to the $\Gamma$ point, we treat $\mathcal{O}(\rho^{3})$ terms as a perturbation. 
%
%
The transition matrix elements of the shift and injection
photocurrents can be readily calculated 
(see SI for details
and the complete list of photoconductivity components):
\begin{align}
       \sigma^{yyy}(\mathbf{k}) = &\dfrac{3 \beta \cos^{2}{\left(\phi \right)}}{\alpha \rho}, \label{eq:I} \\
    \eta^{xxy}(\mathbf{k}) = & \dfrac{ - 2 \alpha \beta\cos{\phi } \cos{3 \phi }}{{4  \rho^{2} \left(\alpha^{2} + 4 \alpha \gamma \rho^{2}\right)}}. \label{eq:eta}
\end{align}
Note that only $\sigma^{yyy}$ yields a finite
response whereas the rest integrate to zero in $\mathbf{k}$-space,
in accordance with Table~\ref{tab:sym}.
Besides predicting a simple polar distribution
for all contributions, Eqs. \eqref{eq:I}-\eqref{eq:eta} 
clearly highlight the importance
of the warping term $\beta$
for a finite quadratic response. This is in line with Refs.~\cite{PhysRevB.95.035134,PhysRevB.107.L161403}, which employed 
the model in the context of topological insulators.

Having described the basic features 
by means of the Rashba model, we now move to analyze 
quantitative results. To this end, we 
have performed fully-relativistic DFT calculations for Au(111) using the Quantum ESPRESSO package \cite{Giannozzi_2009, Giannozzi_2017} 
and computed the nonlinear 
optical properties by means of a  Wannier 
interpolation technique~\cite{PhysRevB.97.245143,PhysRevB.107.205204} (see SI for details).
Fig.~\ref{fig:kslice} presents a \textbf{k}-resolved map of the calculated transition probabilities around $\Gamma$, along with the Rashba model predictions 
overlaid for comparison. 
In the case of  injection [Figs.~\ref{fig:kslice}c,d], 
the \textit{ab initio} results match remarkably well
the model predictions, both in 
the multi-lobe 
pattern form
as well as in the sign.  
In comparison, the Rashba model performs worse in the qualitative description of the shift photocurrent, as evidenced by Figs.~\ref{fig:kslice}a,b. This difference is likely due to the sensitivity of the shift mechanism to the quantum-geometric properties of the Bloch wavefunction 
entering in the expression of $\sigma^{abc}$~\cite{PhysRevB.61.5337}, which are not captured completely in two-band models~\cite{ibanez2022assessing}. 

\begin{figure*}
    \centering
    \includegraphics[width=\textwidth]{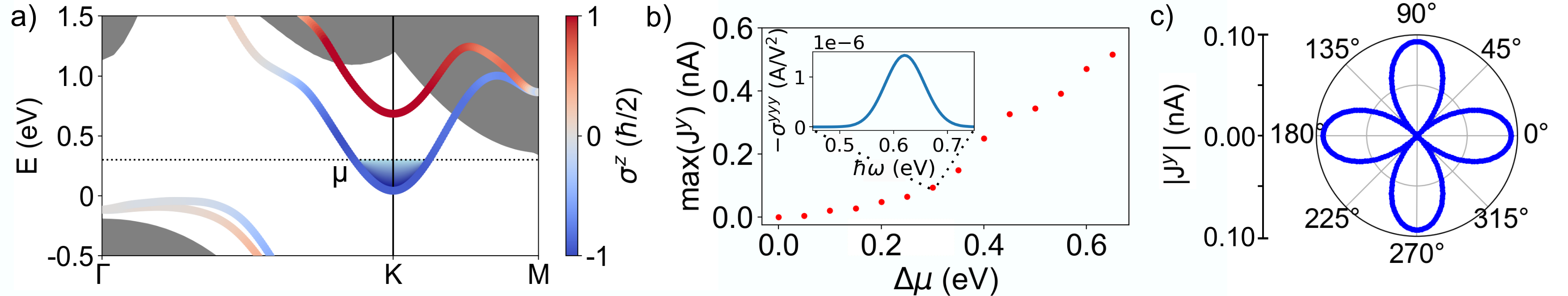}
    \caption{(a) Band structure of Tl/Si(111) showing the out-of-plane spin polarization of the surface bands (given by the color code). The dotted line represents the chemical potential, $\mu$, for a n-doped system. 
    (b) Maximum value of the shift current for different values of the doping. At each value of $\mu$, the frequency and polarization of the light is chosen to maximize the current. 
    The inset shows the 3D shift photoconductivity spectra for $\mu = 0.3$ eV.
    (c) Shift current generated by linearly polarized light with varying polarization angle $\theta$ and frequency $\omega = 0.6$ eV for $\mu=0.3$ eV.}
    \label{fig:Tl}
\end{figure*}

%
Our results for Au(111) highlight the
fundamental features of  quadratic photoresponses
driven by Rashba surface states, 
and open up the possibility 
of extracting the hexagonal warping 
constant $\beta$ through optical measurements of the symmetry-allowed shift 
component $\sigma^{yyy}$. 
However, the metallic nature of bulk Au 
states might present a practical challenge for such purpose due to hybridization with the surface Rashba bands. 
Consequently, we now shift our focus to the Tl/Si(111) surface, whose electronic properties are also prominently influenced by the spin-orbit interaction and possesses important advantages over Au(111).
Firstly, due to the the material's semiconducting nature, the surface response remains well-isolated and distinguishable from any bulk current contribution~\cite{PhysRevLett.102.096805}.
Secondly, it hosts a 
pair of unoccupied spin-split surface states, as shown in 
Fig \ref{fig:Tl},  which in experiment can be 
populated at the 
K, K' valleys under slight electron doping  
without affecting bulk states, thus being extremely robust~\cite{PhysRevMaterials.1.064604}. 
Thirdly, these surface states exhibit 
a huge spin-orbit splitting of $\sim 0.6$ eV 
— a value well-suited for optical applications — accompanied by 
an out-of-plane spin polarization at  the
K and K' valleys, completely deviating from
the Rashba model properties~\cite{PhysRevB.80.241304,PhysRevB.84.125435}.

We have performed \textit{ab-initio} calculations for the Tl/Si(111) surface  following the same procedure as described earlier for Au(111)
(see SI for computational details).
As Tl/Si(111) belongs to the $3m$ point group, the allowed photoconductivity  components are also contained in Table~\ref{tab:sym}.
We therefore focus on the symmetry-allowed shift component
$\sigma^{yyy}$ associated to transitions
between the spin-orbit split surface states indicated in Fig.~\ref{fig:Tl}a, where the lowest band can be populated by electron doping.
%
%

The inset in Fig.~\ref{fig:Tl}b shows that for a moderate doping level of 
$\mu=0.3$ eV, the shift current is largest around $\omega\simeq 0.6$~eV  with maximum 
$|\sigma^{yyy}|\simeq 10^{-6}$ A/V$^{2}$.
Notably, as a central result of this manuscript, the surface-state photoconductivity translates 
into a measurable in-plane photocurrent.
This is proven in main Fig.~\ref{fig:Tl}b, where 
we illustrate how the maximum of $J^{y}$ 
at photon frequency $\omega=0.6$ eV
behaves as a function of doping level.
The calculations account for relevant 
factors such as reflection, absorption 
based on the Glass coefficient~\cite{glass-apl74,tan-cm16}, and the parameters of 
the external  optical field, which were selected to 
align with those commonly used in current experiments~\cite{Ma2019}
(see SI). 
At doping levels ranging from 0 to $\sim$$0.6$ eV, we observe that the 
nonlinear shift photocurrent
flowing along the surface plane is large, comparable to values  reported in prototypic bulk materials like~\cite{measurement3}.
%
%
%
In addition to the magnitude, in Fig.~\ref{fig:Tl}c
we analyze 
the angular dependence of the photocurrent
as a function of the light polarization in
the $xy$-plane. 
We find that the current is maximized along the crystallographic $x$ and $y$ axes, whereas it vanishes at $45 + n\cdot90$~$\deg$ angles. 
This provides a clear signature, different from any potential
linear contribution, 
for experimental detection of pure surface-state shift photocurrents.

In the remaining part of the manuscript, we extend our study to the \emph{spin} photocurrent contribution, thereby  completing the description of surface-state-driven effects.
Nonlinear spin photocurrents have received significantly less attention
than the charge counterparts, with notable progress emerging only recently \cite{PhysRevB.95.035134, Xu2021, PhysRevB.105.045201}. In this work, we adopt the formalism established by Lihm and Park \cite{PhysRevB.105.045201}, 
where transition matrix elements associated to the spin shift and injection currents are naturally suited for Wannier interpolation techniques (see SI).

\begin{figure*}
    \centering
    \includegraphics[width=0.95\textwidth]{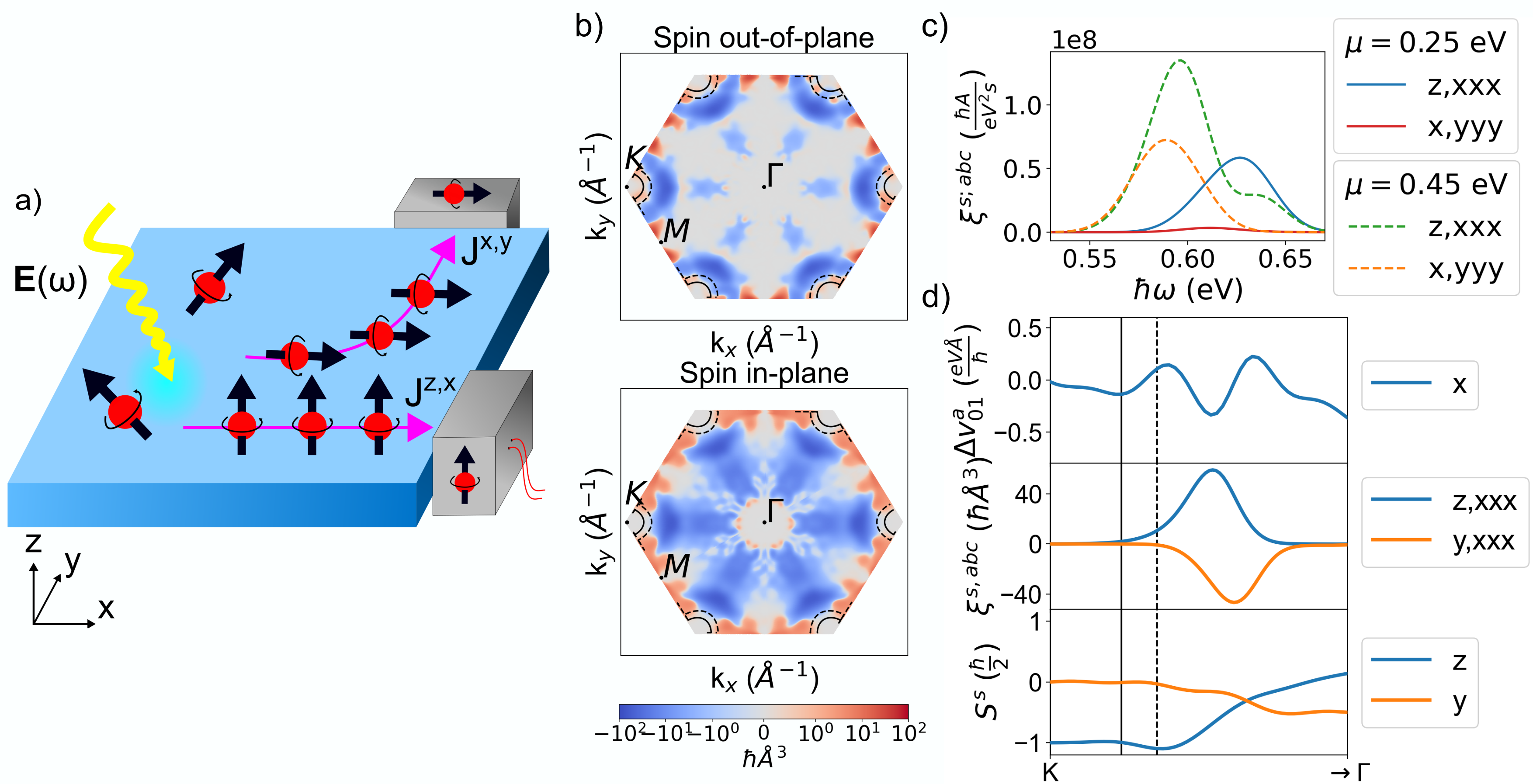}
    \caption{(a) Schematic representation of nonlinear spin current generation by surface states. Out-of-plane ($\hat{\mathbf{z}}$) or in-plane ($\hat{\mathbf{x}}$) spin DC currents can be directed to different contacts at the surface depending on the polarization of the incident light field.
    (b) Brillouin zone \textbf{k}-resolved spin out-of-plane (top) and in-plane (bottom) combinations of the 
    spin injection conductivity tensor components. These combinations are invariant under the C\textsubscript{3} and $M_x$ crystal symmetries of the Tl/Si(111) surface (see SI for additional information). The continuous (dashed) arcs around the K, K' valleys denote the projected Fermi surfaces at $\mu = 0.25$ eV and $\mu = 0.45$ eV.
    (c) Integrated integrated $\xi^{z,xxx}$ and $\xi^{x,yyy}$ components of the spin injection tensor for two representative values of the chemical potential, $\mu = 0.25$ eV and $\mu = 0.45$ eV. Note that at $\mu = 0.25$ eV only the out-of-plane component contributes significantly.
    (d) Top: variation in the difference of the intraband velocity operator along the $\hat{a} = \hat{\mathbf{x}}$ direction shown along the high symmetry $\text{K}-\Gamma$ path. Middle: \textbf{k}-resolved $\xi^{z,xxx}$ and $\xi^{y,xxx}$ spin injection components polarized out-of-plane ($\hat{\mathbf{z}}$, blue curve) and in-plane ($\hat{\mathbf{x}}$, orange curve) along the same path. Bottom: corresponding spin polarization along the same path.
    Vertical solid and dashed lines correspond to the intersection of the path with the Fermi surfaces illustrated in b).
    }
    \label{fig:Tlspin}
\end{figure*}

Table~\ref{tab:sym} lists the symmetry-allowed spin shift and injection
tensor components for Au(111) and Tl/Si(111). 
As sketched in Fig.~\ref{fig:Tlspin}a,
these mechanism provide a remarkable spin-filtering functionality;
they can generate a DC spin current that is spin-polarized along the out-of-plane direction ($z;abc$ components) or in-plane directions depending on the polarization of light.
From the standpoint of symmetry alone, both Au(111) and Tl/Si(111)
possess such functionality.
In practice, however, our DFT calculations 
show that Au(111) generates negligible
out-of-plane 
spin current, which is a direct consequence of the purely in-plane spin texture 
of its surface states (see SI).
%

%
%

In contrast, DFT calculations show that Tl/Si(111) presents both in-plane and out-of-plane spin-polarized photocurrents. 
As surface states are spin-polarized in 
\textbf{k}-space near the band edge, it is particularly relevant to  analyze how the spin texture influences both the form and magnitude of the nonlinear spin photoconductivity.
%
%
%
%
%
%
%
%
%
In Fig.~\ref{fig:Tlspin}b we present the $\mathbf{k}$-resolved contributions for both in-plane and out-of-plane spin injection components, symmetrized to preserve the relevant crystal symmetries.
For small doping levels, we find that the out-of-plane components within the K, K' valleys (top panel) are substantially larger than the in-plane components (bottom panel).
The dominant contribution arises through the component $\xi^{z,xxx}$, while in-plane components such as $\xi^{x,yyy}$ 
are significantly smaller. The latter  become  larger in other regions of the Brillouin zone, but are energetically accessible only 
for far larger doping levels $\mu > 0.6$ eV.
Thus, in practice, Tl/Si(111) acts as a spin injector/filter
as a function of doping, 
generating spin-polarized currents with a well-defined out-of-plane spin component ($S_z$) for $\mu < 0.6$ eV. 
This is rather remarkable, given that the spin is not
a good quantum number of 
this system due to its strong relativistic effects.

%

To further illustrate this feature of the Tl/Si(111) surface, Fig. \ref{fig:Tlspin}c presents the dominant integrated components of the spin injection tensor $\xi^{z,xxx}$ and $\xi^{x,yyy}$, for two representative values of the chemical potential, $\mu = 0.25$ and $\mu = 0.45$ eV. At $\mu = 0.25$ eV, the spin injection is predominantly governed by the out-of-plane component, whereas at $\mu = 0.45$ eV, the in-plane contribution becomes quantitatively relevant. 
In order to understand the origin of these results,
in Fig.~\ref{fig:Tlspin}d we display the k-resolved 
structure of several relevant quantities along the high-symmetry path 
$\text{K}-\Gamma$, where we mark the Fermi surface cuts
for the two doping levels $\mu = 0.25$ and 0.45 eV. 
We begin by considering the top panel, which displays the
difference of velocity matrix elements entering the
expression of the spin injection contribution (see SI).
Since both surface states exhibit a band minimum at K (see Fig.~\ref{fig:Tl}a), this quantity remains very small near this valley.
As a result, 
the two 
spin injection contributions $z,xxx$ and $y,xxx$, 
shown in the middle panel of Fig.~\ref{fig:Tlspin}d are also very small in this region, with both peaking away from K.
Note that these components are not the same as in Fig.~\ref{fig:Tlspin}c, the ones considered in Fig.~\ref{fig:Tlspin}d differ only in the spin index for comparative purposes.
Notably, the origin of the strong out-of-plane
$z,xxx$ peak
can be traced back to the out-of-plane spin-polarization
of the surface states, illustrated in the lower panel of Fig.~\ref{fig:Tlspin}d. 
A comparison with the middle panel reveals that the $z,xxx$ spin injection component peaks in the region where $S_z$ dominates, whereas the $y,xxx$ component peaks only when $S_z$ becomes significantly smaller. This highlights the crucial role of the out-of-plane spin polarization -- completely absent in Au(111) -- in generating a pure out-of-plane nonlinear spin current in Tl/Si(111).

In summary, we have presented a detailed study of nonlinear photocurrents arising from spin-orbit split surface states.
Besides elucidating the fundamental aspects of the effect in the 
Rashba prototype Au(111)
surface, we have identified 
an optimal platform for experimental detection in the Tl/Si(111) surface. 
Our calculations have shown that the magnitude of the generated bulk photovoltaic
effect is well within the measurable limits, and we have provided clear angular fingerprints for
its detection. 
While we have focused on Tl/Si(111), our proposal is general given that it does not require inversion-symmetry breaking of the bulk substrate, hence we expect similar features in other heavily studied compounds made of Bi, Pb or Ag~\cite{bihlmayer_magnetic_nodate,PhysRevB.75.195414}. 
Additionally, we have explored the nonlinear spin counterpart of this effect, demonstrating that 
Tl/Si(111) functions as an exceptionally effective spin filter, with its performance tunable through doping.
Notably, nonlinear spin currents provide far greater flexibility than the conventional linear approach, which is limited by the lower dimensionality of the spin Hall conductivity tensor and only permits transverse spin accumulation~\cite{SPHmeassurement, Sinova2004, Schliemann2006}.
The interplay between spin texture and nonlinear responses paves the way for novel strategies in engineering spatial- and spin-selective nonlinear spin generation, offering enhanced control and tunability for spintronic applications.

%
%

%
%
%


J. S., P. G.-G. and J. I. A. acknowledge the financial support of the European Union’s Horizon 2020 research and innovation programme under the European Research Council (ERC) grant agreement No. 946629 StG PhotoNow. 
D. H.-P. is grateful for funding from the Diputaci\'on Foral de Gipuzkoa through Grants 2023-FELL-000002-01, 2024-FELL-000009-01. D.H.-P. and J. I.-A. are also grateful from support of the Spanish {MICIU/AEI/10.13039/501100011033} through Project No. PID2023-147324NA-I00.


\bibliography{bibliography}

\begin{thebibliography}{44}%
\makeatletter
\providecommand \@ifxundefined [1]{%
 \@ifx{#1\undefined}
}%
\providecommand \@ifnum [1]{%
 \ifnum #1\expandafter \@firstoftwo
 \else \expandafter \@secondoftwo
 \fi
}%
\providecommand \@ifx [1]{%
 \ifx #1\expandafter \@firstoftwo
 \else \expandafter \@secondoftwo
 \fi
}%
\providecommand \natexlab [1]{#1}%
\providecommand \enquote  [1]{``#1''}%
\providecommand \bibnamefont  [1]{#1}%
\providecommand \bibfnamefont [1]{#1}%
\providecommand \citenamefont [1]{#1}%
\providecommand \href@noop [0]{\@secondoftwo}%
\providecommand \href [0]{\begingroup \@sanitize@url \@href}%
\providecommand \@href[1]{\@@startlink{#1}\@@href}%
\providecommand \@@href[1]{\endgroup#1\@@endlink}%
\providecommand \@sanitize@url [0]{\catcode `\\12\catcode `\$12\catcode `\&12\catcode `\#12\catcode `\^12\catcode `\_12\catcode `\%12\relax}%
\providecommand \@@startlink[1]{}%
\providecommand \@@endlink[0]{}%
\providecommand \url  [0]{\begingroup\@sanitize@url \@url }%
\providecommand \@url [1]{\endgroup\@href {#1}{\urlprefix }}%
\providecommand \urlprefix  [0]{URL }%
\providecommand \Eprint [0]{\href }%
\providecommand \doibase [0]{https://doi.org/}%
\providecommand \selectlanguage [0]{\@gobble}%
\providecommand \bibinfo  [0]{\@secondoftwo}%
\providecommand \bibfield  [0]{\@secondoftwo}%
\providecommand \translation [1]{[#1]}%
\providecommand \BibitemOpen [0]{}%
\providecommand \bibitemStop [0]{}%
\providecommand \bibitemNoStop [0]{.\EOS\space}%
\providecommand \EOS [0]{\spacefactor3000\relax}%
\providecommand \BibitemShut  [1]{\csname bibitem#1\endcsname}%
\let\auto@bib@innerbib\@empty
\bibitem [{\citenamefont {Belinicher}\ \emph {et~al.}(1982)\citenamefont {Belinicher}, \citenamefont {Ivchenko},\ and\ \citenamefont {Sturman}}]{belinicher-jetp82}%
  \BibitemOpen
  \bibfield  {author} {\bibinfo {author} {\bibfnamefont {V.~I.}\ \bibnamefont {Belinicher}}, \bibinfo {author} {\bibfnamefont {E.~L.}\ \bibnamefont {Ivchenko}},\ and\ \bibinfo {author} {\bibfnamefont {B.~I.}\ \bibnamefont {Sturman}},\ }\bibfield  {title} {\bibinfo {title} {Kinetic theory of the displacement photovoltaic effect in piezoelectrics},\ }\href {http://www.jetp.ac.ru/cgi-bin/e/index/e/56/2/p359?a=list} {\bibfield  {journal} {\bibinfo  {journal} {Sov. Phys. JETP}\ }\textbf {\bibinfo {volume} {56}},\ \bibinfo {pages} {359} (\bibinfo {year} {1982})}\BibitemShut {NoStop}%
\bibitem [{\citenamefont {Sturman}\ and\ \citenamefont {Fridkin}(1992)}]{sturman-book92}%
  \BibitemOpen
  \bibfield  {author} {\bibinfo {author} {\bibfnamefont {B.~I.}\ \bibnamefont {Sturman}}\ and\ \bibinfo {author} {\bibfnamefont {V.~M.}\ \bibnamefont {Fridkin}},\ }\href@noop {} {\emph {\bibinfo {title} {The photovoltaic and photorefractive effects in noncentrosymmetric materials}}}\ (\bibinfo  {publisher} {Gordon and Breach},\ \bibinfo {year} {1992})\BibitemShut {NoStop}%
\bibitem [{\citenamefont {von Baltz}\ and\ \citenamefont {Kraut}(1981)}]{baltz-prb81}%
  \BibitemOpen
  \bibfield  {author} {\bibinfo {author} {\bibfnamefont {R.}~\bibnamefont {von Baltz}}\ and\ \bibinfo {author} {\bibfnamefont {W.}~\bibnamefont {Kraut}},\ }\bibfield  {title} {\bibinfo {title} {Theory of the bulk photovoltaic effect in pure crystals},\ }\href {https://doi.org/10.1103/PhysRevB.23.5590} {\bibfield  {journal} {\bibinfo  {journal} {Phys. Rev. B}\ }\textbf {\bibinfo {volume} {23}},\ \bibinfo {pages} {5590} (\bibinfo {year} {1981})}\BibitemShut {NoStop}%
\bibitem [{\citenamefont {Ma}\ \emph {et~al.}(2019)\citenamefont {Ma}, \citenamefont {Gu}, \citenamefont {Liu}, \citenamefont {Lai}, \citenamefont {Yu}, \citenamefont {Zhuo}, \citenamefont {Liu}, \citenamefont {Chen}, \citenamefont {Feng},\ and\ \citenamefont {Sun}}]{Ma2019}%
  \BibitemOpen
  \bibfield  {author} {\bibinfo {author} {\bibfnamefont {J.}~\bibnamefont {Ma}}, \bibinfo {author} {\bibfnamefont {Q.}~\bibnamefont {Gu}}, \bibinfo {author} {\bibfnamefont {Y.}~\bibnamefont {Liu}}, \bibinfo {author} {\bibfnamefont {J.}~\bibnamefont {Lai}}, \bibinfo {author} {\bibfnamefont {P.}~\bibnamefont {Yu}}, \bibinfo {author} {\bibfnamefont {X.}~\bibnamefont {Zhuo}}, \bibinfo {author} {\bibfnamefont {Z.}~\bibnamefont {Liu}}, \bibinfo {author} {\bibfnamefont {J.-H.}\ \bibnamefont {Chen}}, \bibinfo {author} {\bibfnamefont {J.}~\bibnamefont {Feng}},\ and\ \bibinfo {author} {\bibfnamefont {D.}~\bibnamefont {Sun}},\ }\bibfield  {title} {\bibinfo {title} {Nonlinear photoresponse of type-{II} {W}eyl semimetals},\ }\href {https://doi.org/10.1038/s41563-019-0296-5} {\bibfield  {journal} {\bibinfo  {journal} {Nature Materials}\ }\textbf {\bibinfo {volume} {18}},\ \bibinfo {pages} {476} (\bibinfo {year} {2019})}\BibitemShut {NoStop}%
\bibitem [{\citenamefont {Osterhoudt}\ \emph {et~al.}(2019)\citenamefont {Osterhoudt}, \citenamefont {Diebel}, \citenamefont {Gray}, \citenamefont {Yang}, \citenamefont {Stanco}, \citenamefont {Huang}, \citenamefont {Shen}, \citenamefont {Ni}, \citenamefont {Moll}, \citenamefont {Ran} \emph {et~al.}}]{osterhoudt2019colossal}%
  \BibitemOpen
  \bibfield  {author} {\bibinfo {author} {\bibfnamefont {G.~B.}\ \bibnamefont {Osterhoudt}}, \bibinfo {author} {\bibfnamefont {L.~K.}\ \bibnamefont {Diebel}}, \bibinfo {author} {\bibfnamefont {M.~J.}\ \bibnamefont {Gray}}, \bibinfo {author} {\bibfnamefont {X.}~\bibnamefont {Yang}}, \bibinfo {author} {\bibfnamefont {J.}~\bibnamefont {Stanco}}, \bibinfo {author} {\bibfnamefont {X.}~\bibnamefont {Huang}}, \bibinfo {author} {\bibfnamefont {B.}~\bibnamefont {Shen}}, \bibinfo {author} {\bibfnamefont {N.}~\bibnamefont {Ni}}, \bibinfo {author} {\bibfnamefont {P.~J.}\ \bibnamefont {Moll}}, \bibinfo {author} {\bibfnamefont {Y.}~\bibnamefont {Ran}}, \emph {et~al.},\ }\bibfield  {title} {\bibinfo {title} {Colossal mid-infrared bulk photovoltaic effect in a type-{I} {W}eyl semimetal},\ }\href {https://www.nature.com/articles/s41563-019-0297-4} {\bibfield  {journal} {\bibinfo  {journal} {Nature Materials}\ }\textbf {\bibinfo {volume} {18}},\ \bibinfo {pages} {471} (\bibinfo {year} {2019})}\BibitemShut {NoStop}%
\bibitem [{\citenamefont {Zhang}\ \emph {et~al.}(2019)\citenamefont {Zhang}, \citenamefont {Ideue}, \citenamefont {Onga}, \citenamefont {Qin}, \citenamefont {Suzuki}, \citenamefont {Zak}, \citenamefont {Tenne}, \citenamefont {Smet},\ and\ \citenamefont {Iwasa}}]{zhang_enhanced_2019}%
  \BibitemOpen
  \bibfield  {author} {\bibinfo {author} {\bibfnamefont {Y.~J.}\ \bibnamefont {Zhang}}, \bibinfo {author} {\bibfnamefont {T.}~\bibnamefont {Ideue}}, \bibinfo {author} {\bibfnamefont {M.}~\bibnamefont {Onga}}, \bibinfo {author} {\bibfnamefont {F.}~\bibnamefont {Qin}}, \bibinfo {author} {\bibfnamefont {R.}~\bibnamefont {Suzuki}}, \bibinfo {author} {\bibfnamefont {A.}~\bibnamefont {Zak}}, \bibinfo {author} {\bibfnamefont {R.}~\bibnamefont {Tenne}}, \bibinfo {author} {\bibfnamefont {J.~H.}\ \bibnamefont {Smet}},\ and\ \bibinfo {author} {\bibfnamefont {Y.}~\bibnamefont {Iwasa}},\ }\bibfield  {title} {\bibinfo {title} {Enhanced intrinsic photovoltaic effect in tungsten disulfide nanotubes},\ }\href {https://doi.org/10.1038/s41586-019-1303-3} {\bibfield  {journal} {\bibinfo  {journal} {Nature}\ }\textbf {\bibinfo {volume} {570}},\ \bibinfo {pages} {349} (\bibinfo {year} {2019})}\BibitemShut {NoStop}%
\bibitem [{\citenamefont {Krishna}\ \emph {et~al.}(2023)\citenamefont {Krishna}, \citenamefont {Garcia-Goiricelaya}, \citenamefont {de~Juan},\ and\ \citenamefont {Iba\~nez{-}Azpiroz}}]{PhysRevB.108.165418}%
  \BibitemOpen
  \bibfield  {author} {\bibinfo {author} {\bibfnamefont {J.}~\bibnamefont {Krishna}}, \bibinfo {author} {\bibfnamefont {P.}~\bibnamefont {Garcia-Goiricelaya}}, \bibinfo {author} {\bibfnamefont {F.}~\bibnamefont {de~Juan}},\ and\ \bibinfo {author} {\bibfnamefont {J.}~\bibnamefont {Iba\~nez{-}Azpiroz}},\ }\bibfield  {title} {\bibinfo {title} {Understanding the large shift photocurrent of {WS}\textsubscript{2} nanotubes: A comparative analysis with monolayers},\ }\href {https://doi.org/10.1103/PhysRevB.108.165418} {\bibfield  {journal} {\bibinfo  {journal} {Phys. Rev. B}\ }\textbf {\bibinfo {volume} {108}},\ \bibinfo {pages} {165418} (\bibinfo {year} {2023})}\BibitemShut {NoStop}%
\bibitem [{\citenamefont {Lihm}\ and\ \citenamefont {Park}(2022)}]{PhysRevB.105.045201}%
  \BibitemOpen
  \bibfield  {author} {\bibinfo {author} {\bibfnamefont {J.-M.}\ \bibnamefont {Lihm}}\ and\ \bibinfo {author} {\bibfnamefont {C.-H.}\ \bibnamefont {Park}},\ }\bibfield  {title} {\bibinfo {title} {Comprehensive theory of second-order spin photocurrents},\ }\href {https://doi.org/10.1103/PhysRevB.105.045201} {\bibfield  {journal} {\bibinfo  {journal} {Phys. Rev. B}\ }\textbf {\bibinfo {volume} {105}},\ \bibinfo {pages} {045201} (\bibinfo {year} {2022})}\BibitemShut {NoStop}%
\bibitem [{\citenamefont {Xu}\ \emph {et~al.}(2021{\natexlab{a}})\citenamefont {Xu}, \citenamefont {Wang}, \citenamefont {Zhou},\ and\ \citenamefont {Li}}]{xu_pure_2021}%
  \BibitemOpen
  \bibfield  {author} {\bibinfo {author} {\bibfnamefont {H.}~\bibnamefont {Xu}}, \bibinfo {author} {\bibfnamefont {H.}~\bibnamefont {Wang}}, \bibinfo {author} {\bibfnamefont {J.}~\bibnamefont {Zhou}},\ and\ \bibinfo {author} {\bibfnamefont {J.}~\bibnamefont {Li}},\ }\bibfield  {title} {\bibinfo {title} {Pure spin photocurrent in non-centrosymmetric crystals: bulk spin photovoltaic effect},\ }\href {https://doi.org/10.1038/s41467-021-24541-7} {\bibfield  {journal} {\bibinfo  {journal} {Nature Communications}\ }\textbf {\bibinfo {volume} {12}},\ \bibinfo {pages} {4330} (\bibinfo {year} {2021}{\natexlab{a}})},\ \bibinfo {note} {publisher: Nature Publishing Group}\BibitemShut {NoStop}%
\bibitem [{\citenamefont {Dong}\ \emph {et~al.}(2024)\citenamefont {Dong}, \citenamefont {Cao}, \citenamefont {Tan},\ and\ \citenamefont {Fei}}]{dong2024crystalsymmetryselectedpure}%
  \BibitemOpen
  \bibfield  {author} {\bibinfo {author} {\bibfnamefont {R.}~\bibnamefont {Dong}}, \bibinfo {author} {\bibfnamefont {R.}~\bibnamefont {Cao}}, \bibinfo {author} {\bibfnamefont {D.}~\bibnamefont {Tan}},\ and\ \bibinfo {author} {\bibfnamefont {R.}~\bibnamefont {Fei}},\ }\href {https://arxiv.org/abs/2412.09216} {\bibinfo {title} {Crystal symmetry selected pure spin photocurrent in altermagnetic insulators}} (\bibinfo {year} {2024}),\ \Eprint {https://arxiv.org/abs/2412.09216} {arXiv:2412.09216 [cond-mat.mtrl-sci]} \BibitemShut {NoStop}%
\bibitem [{\citenamefont {Arzate}\ \emph {et~al.}(2014)\citenamefont {Arzate}, \citenamefont {V\'azquez-Nava},\ and\ \citenamefont {Mendoza}}]{PhysRevB.90.205310}%
  \BibitemOpen
  \bibfield  {author} {\bibinfo {author} {\bibfnamefont {N.}~\bibnamefont {Arzate}}, \bibinfo {author} {\bibfnamefont {R.~A.}\ \bibnamefont {V\'azquez-Nava}},\ and\ \bibinfo {author} {\bibfnamefont {B.~S.}\ \bibnamefont {Mendoza}},\ }\bibfield  {title} {\bibinfo {title} {Optical spin- and current-injection study on {Si(111)-In} surfaces},\ }\href {https://doi.org/10.1103/PhysRevB.90.205310} {\bibfield  {journal} {\bibinfo  {journal} {Phys. Rev. B}\ }\textbf {\bibinfo {volume} {90}},\ \bibinfo {pages} {205310} (\bibinfo {year} {2014})}\BibitemShut {NoStop}%
\bibitem [{\citenamefont {Cabellos}\ \emph {et~al.}(2011)\citenamefont {Cabellos}, \citenamefont {Mendoza},\ and\ \citenamefont {Shkrebtii}}]{PhysRevB.84.195326}%
  \BibitemOpen
  \bibfield  {author} {\bibinfo {author} {\bibfnamefont {J.~L.}\ \bibnamefont {Cabellos}}, \bibinfo {author} {\bibfnamefont {B.~S.}\ \bibnamefont {Mendoza}},\ and\ \bibinfo {author} {\bibfnamefont {A.~I.}\ \bibnamefont {Shkrebtii}},\ }\bibfield  {title} {\bibinfo {title} {Optical coherent current control at surfaces: Theory of injection current},\ }\href {https://doi.org/10.1103/PhysRevB.84.195326} {\bibfield  {journal} {\bibinfo  {journal} {Phys. Rev. B}\ }\textbf {\bibinfo {volume} {84}},\ \bibinfo {pages} {195326} (\bibinfo {year} {2011})}\BibitemShut {NoStop}%
\bibitem [{\citenamefont {Kim}\ \emph {et~al.}(2017)\citenamefont {Kim}, \citenamefont {Morimoto},\ and\ \citenamefont {Nagaosa}}]{PhysRevB.95.035134}%
  \BibitemOpen
  \bibfield  {author} {\bibinfo {author} {\bibfnamefont {K.~W.}\ \bibnamefont {Kim}}, \bibinfo {author} {\bibfnamefont {T.}~\bibnamefont {Morimoto}},\ and\ \bibinfo {author} {\bibfnamefont {N.}~\bibnamefont {Nagaosa}},\ }\bibfield  {title} {\bibinfo {title} {Shift charge and spin photocurrents in {D}irac surface states of topological insulator},\ }\href {https://doi.org/10.1103/PhysRevB.95.035134} {\bibfield  {journal} {\bibinfo  {journal} {Phys. Rev. B}\ }\textbf {\bibinfo {volume} {95}},\ \bibinfo {pages} {035134} (\bibinfo {year} {2017})}\BibitemShut {NoStop}%
\bibitem [{\citenamefont {Leppenen}\ and\ \citenamefont {Golub}(2023)}]{PhysRevB.107.L161403}%
  \BibitemOpen
  \bibfield  {author} {\bibinfo {author} {\bibfnamefont {N.~V.}\ \bibnamefont {Leppenen}}\ and\ \bibinfo {author} {\bibfnamefont {L.~E.}\ \bibnamefont {Golub}},\ }\bibfield  {title} {\bibinfo {title} {Linear photogalvanic effect in surface states of topological insulators},\ }\href {https://doi.org/10.1103/PhysRevB.107.L161403} {\bibfield  {journal} {\bibinfo  {journal} {Phys. Rev. B}\ }\textbf {\bibinfo {volume} {107}},\ \bibinfo {pages} {L161403} (\bibinfo {year} {2023})}\BibitemShut {NoStop}%
\bibitem [{\citenamefont {Sipe}\ and\ \citenamefont {Shkrebtii}(2000)}]{PhysRevB.61.5337}%
  \BibitemOpen
  \bibfield  {author} {\bibinfo {author} {\bibfnamefont {J.~E.}\ \bibnamefont {Sipe}}\ and\ \bibinfo {author} {\bibfnamefont {A.~I.}\ \bibnamefont {Shkrebtii}},\ }\bibfield  {title} {\bibinfo {title} {Second-order optical response in semiconductors},\ }\href {https://doi.org/10.1103/PhysRevB.61.5337} {\bibfield  {journal} {\bibinfo  {journal} {Phys. Rev. B}\ }\textbf {\bibinfo {volume} {61}},\ \bibinfo {pages} {5337} (\bibinfo {year} {2000})}\BibitemShut {NoStop}%
\bibitem [{sup()}]{sup}%
  \BibitemOpen
  \bibfield  {title} {\bibinfo {title} {See supplementary material for additional information of the charge and spin photonconductivity expression},\ }\href@noop {} {\bibinfo  {journal} {model and material calculations}\ }\BibitemShut {NoStop}%
\bibitem [{\citenamefont {LaShell}\ \emph {et~al.}(1996)\citenamefont {LaShell}, \citenamefont {McDougall},\ and\ \citenamefont {Jensen}}]{PhysRevLett.77.3419}%
  \BibitemOpen
\bibfield  {journal} {  }\bibfield  {author} {\bibinfo {author} {\bibfnamefont {S.}~\bibnamefont {LaShell}}, \bibinfo {author} {\bibfnamefont {B.~A.}\ \bibnamefont {McDougall}},\ and\ \bibinfo {author} {\bibfnamefont {E.}~\bibnamefont {Jensen}},\ }\bibfield  {title} {\bibinfo {title} {Spin splitting of an {Au(111)} surface state band observed with angle resolved photoelectron spectroscopy},\ }\href {https://doi.org/10.1103/PhysRevLett.77.3419} {\bibfield  {journal} {\bibinfo  {journal} {Phys. Rev. Lett.}\ }\textbf {\bibinfo {volume} {77}},\ \bibinfo {pages} {3419} (\bibinfo {year} {1996})}\BibitemShut {NoStop}%
\bibitem [{\citenamefont {Heide}\ \emph {et~al.}(2006)\citenamefont {Heide}, \citenamefont {Bihlmayer}, \citenamefont {Mavropoulos}, \citenamefont {Bringer},\ and\ \citenamefont {Bl\"{u}gel}}]{sobihlmayer}%
  \BibitemOpen
  \bibfield  {author} {\bibinfo {author} {\bibfnamefont {M.}~\bibnamefont {Heide}}, \bibinfo {author} {\bibfnamefont {G.}~\bibnamefont {Bihlmayer}}, \bibinfo {author} {\bibfnamefont {P.}~\bibnamefont {Mavropoulos}}, \bibinfo {author} {\bibfnamefont {A.}~\bibnamefont {Bringer}},\ and\ \bibinfo {author} {\bibfnamefont {S.}~\bibnamefont {Bl\"{u}gel}},\ }\href@noop {} {\bibfield  {journal} {\bibinfo  {journal} {Psi-k Newsletter}\ }\textbf {\bibinfo {volume} {78}},\ \bibinfo {pages} {1109} (\bibinfo {year} {2006})}\BibitemShut {NoStop}%
\bibitem [{\citenamefont {Henk}\ \emph {et~al.}(2004)\citenamefont {Henk}, \citenamefont {Hoesch}, \citenamefont {Osterwalder}, \citenamefont {Ernst},\ and\ \citenamefont {Bruno}}]{au111-exp}%
  \BibitemOpen
  \bibfield  {author} {\bibinfo {author} {\bibfnamefont {J.}~\bibnamefont {Henk}}, \bibinfo {author} {\bibfnamefont {M.}~\bibnamefont {Hoesch}}, \bibinfo {author} {\bibfnamefont {J.}~\bibnamefont {Osterwalder}}, \bibinfo {author} {\bibfnamefont {A.}~\bibnamefont {Ernst}},\ and\ \bibinfo {author} {\bibfnamefont {P.}~\bibnamefont {Bruno}},\ }\href@noop {} {\bibfield  {journal} {\bibinfo  {journal} {Journal of Physics: Condensed Matter}\ }\textbf {\bibinfo {volume} {16}},\ \bibinfo {pages} {7581} (\bibinfo {year} {2004})}\BibitemShut {NoStop}%
\bibitem [{\citenamefont {Iba\~nez{-}Azpiroz}\ \emph {et~al.}(2013)\citenamefont {Iba\~nez{-}Azpiroz}, \citenamefont {Bergara}, \citenamefont {Sherman},\ and\ \citenamefont {Eiguren}}]{PhysRevB.88.125404}%
  \BibitemOpen
  \bibfield  {author} {\bibinfo {author} {\bibfnamefont {J.}~\bibnamefont {Iba\~nez{-}Azpiroz}}, \bibinfo {author} {\bibfnamefont {A.}~\bibnamefont {Bergara}}, \bibinfo {author} {\bibfnamefont {E.~Y.}\ \bibnamefont {Sherman}},\ and\ \bibinfo {author} {\bibfnamefont {A.}~\bibnamefont {Eiguren}},\ }\bibfield  {title} {\bibinfo {title} {Spin-flip transitions and departure from the {R}ashba model in the {Au(111)} surface},\ }\href {https://doi.org/10.1103/PhysRevB.88.125404} {\bibfield  {journal} {\bibinfo  {journal} {Phys. Rev. B}\ }\textbf {\bibinfo {volume} {88}},\ \bibinfo {pages} {125404} (\bibinfo {year} {2013})}\BibitemShut {NoStop}%
\bibitem [{\citenamefont {Fu}(2009)}]{PhysRevLett.103.266801}%
  \BibitemOpen
  \bibfield  {author} {\bibinfo {author} {\bibfnamefont {L.}~\bibnamefont {Fu}},\ }\bibfield  {title} {\bibinfo {title} {Hexagonal warping effects in the surface states of the topological insulator {Bi}\textsubscript{2}{Te}\textsubscript{3}},\ }\href {https://doi.org/10.1103/PhysRevLett.103.266801} {\bibfield  {journal} {\bibinfo  {journal} {Phys. Rev. Lett.}\ }\textbf {\bibinfo {volume} {103}},\ \bibinfo {pages} {266801} (\bibinfo {year} {2009})}\BibitemShut {NoStop}%
\bibitem [{\citenamefont {Giannozzi}\ \emph {et~al.}(2009)\citenamefont {Giannozzi}, \citenamefont {Baroni}, \citenamefont {Bonini}, \citenamefont {Calandra}, \citenamefont {Car}, \citenamefont {Cavazzoni}, \citenamefont {Ceresoli}, \citenamefont {Chiarotti}, \citenamefont {Cococcioni}, \citenamefont {Dabo}, \citenamefont {Dal~Corso}, \citenamefont {de~Gironcoli}, \citenamefont {Fabris}, \citenamefont {Fratesi}, \citenamefont {Gebauer}, \citenamefont {Gerstmann}, \citenamefont {Gougoussis}, \citenamefont {Kokalj}, \citenamefont {Lazzeri}, \citenamefont {Martin-Samos}, \citenamefont {Marzari}, \citenamefont {Mauri}, \citenamefont {Mazzarello}, \citenamefont {Paolini}, \citenamefont {Pasquarello}, \citenamefont {Paulatto}, \citenamefont {Sbraccia}, \citenamefont {Scandolo}, \citenamefont {Sclauzero}, \citenamefont {Seitsonen}, \citenamefont {Smogunov}, \citenamefont {Umari},\ and\ \citenamefont {Wentzcovitch}}]{Giannozzi_2009}%
  \BibitemOpen
  \bibfield  {author} {\bibinfo {author} {\bibfnamefont {P.}~\bibnamefont {Giannozzi}}, \bibinfo {author} {\bibfnamefont {S.}~\bibnamefont {Baroni}}, \bibinfo {author} {\bibfnamefont {N.}~\bibnamefont {Bonini}}, \bibinfo {author} {\bibfnamefont {M.}~\bibnamefont {Calandra}}, \bibinfo {author} {\bibfnamefont {R.}~\bibnamefont {Car}}, \bibinfo {author} {\bibfnamefont {C.}~\bibnamefont {Cavazzoni}}, \bibinfo {author} {\bibfnamefont {D.}~\bibnamefont {Ceresoli}}, \bibinfo {author} {\bibfnamefont {G.~L.}\ \bibnamefont {Chiarotti}}, \bibinfo {author} {\bibfnamefont {M.}~\bibnamefont {Cococcioni}}, \bibinfo {author} {\bibfnamefont {I.}~\bibnamefont {Dabo}}, \bibinfo {author} {\bibfnamefont {A.}~\bibnamefont {Dal~Corso}}, \bibinfo {author} {\bibfnamefont {S.}~\bibnamefont {de~Gironcoli}}, \bibinfo {author} {\bibfnamefont {S.}~\bibnamefont {Fabris}}, \bibinfo {author} {\bibfnamefont {G.}~\bibnamefont {Fratesi}}, \bibinfo {author} {\bibfnamefont {R.}~\bibnamefont {Gebauer}}, \bibinfo {author} {\bibfnamefont
  {U.}~\bibnamefont {Gerstmann}}, \bibinfo {author} {\bibfnamefont {C.}~\bibnamefont {Gougoussis}}, \bibinfo {author} {\bibfnamefont {A.}~\bibnamefont {Kokalj}}, \bibinfo {author} {\bibfnamefont {M.}~\bibnamefont {Lazzeri}}, \bibinfo {author} {\bibfnamefont {L.}~\bibnamefont {Martin-Samos}}, \bibinfo {author} {\bibfnamefont {N.}~\bibnamefont {Marzari}}, \bibinfo {author} {\bibfnamefont {F.}~\bibnamefont {Mauri}}, \bibinfo {author} {\bibfnamefont {R.}~\bibnamefont {Mazzarello}}, \bibinfo {author} {\bibfnamefont {S.}~\bibnamefont {Paolini}}, \bibinfo {author} {\bibfnamefont {A.}~\bibnamefont {Pasquarello}}, \bibinfo {author} {\bibfnamefont {L.}~\bibnamefont {Paulatto}}, \bibinfo {author} {\bibfnamefont {C.}~\bibnamefont {Sbraccia}}, \bibinfo {author} {\bibfnamefont {S.}~\bibnamefont {Scandolo}}, \bibinfo {author} {\bibfnamefont {G.}~\bibnamefont {Sclauzero}}, \bibinfo {author} {\bibfnamefont {A.~P.}\ \bibnamefont {Seitsonen}}, \bibinfo {author} {\bibfnamefont {A.}~\bibnamefont {Smogunov}}, \bibinfo {author}
  {\bibfnamefont {P.}~\bibnamefont {Umari}},\ and\ \bibinfo {author} {\bibfnamefont {R.~M.}\ \bibnamefont {Wentzcovitch}},\ }\bibfield  {title} {\bibinfo {title} {{QUANTUM ESPRESSO}: a modular and open-source software project for quantum simulations of materials},\ }\href {https://doi.org/10.1088/0953-8984/21/39/395502} {\bibfield  {journal} {\bibinfo  {journal} {Journal of Physics: Condensed Matter}\ }\textbf {\bibinfo {volume} {21}},\ \bibinfo {pages} {395502} (\bibinfo {year} {2009})}\BibitemShut {NoStop}%
\bibitem [{\citenamefont {Giannozzi}\ \emph {et~al.}(2017)\citenamefont {Giannozzi}, \citenamefont {Andreussi}, \citenamefont {Brumme}, \citenamefont {Bunau}, \citenamefont {Buongiorno~Nardelli}, \citenamefont {Calandra}, \citenamefont {Car}, \citenamefont {Cavazzoni}, \citenamefont {Ceresoli}, \citenamefont {Cococcioni}, \citenamefont {Colonna}, \citenamefont {Carnimeo}, \citenamefont {Dal~Corso}, \citenamefont {de~Gironcoli}, \citenamefont {Delugas}, \citenamefont {DiStasio}, \citenamefont {Ferretti}, \citenamefont {Floris}, \citenamefont {Fratesi}, \citenamefont {Fugallo}, \citenamefont {Gebauer}, \citenamefont {Gerstmann}, \citenamefont {Giustino}, \citenamefont {Gorni}, \citenamefont {Jia}, \citenamefont {Kawamura}, \citenamefont {Ko}, \citenamefont {Kokalj}, \citenamefont {Küçükbenli}, \citenamefont {Lazzeri}, \citenamefont {Marsili}, \citenamefont {Marzari}, \citenamefont {Mauri}, \citenamefont {Nguyen}, \citenamefont {Nguyen}, \citenamefont {Otero-de-la Roza}, \citenamefont {Paulatto},
  \citenamefont {Poncé}, \citenamefont {Rocca}, \citenamefont {Sabatini}, \citenamefont {Santra}, \citenamefont {Schlipf}, \citenamefont {Seitsonen}, \citenamefont {Smogunov}, \citenamefont {Timrov}, \citenamefont {Thonhauser}, \citenamefont {Umari}, \citenamefont {Vast}, \citenamefont {Wu},\ and\ \citenamefont {Baroni}}]{Giannozzi_2017}%
  \BibitemOpen
  \bibfield  {author} {\bibinfo {author} {\bibfnamefont {P.}~\bibnamefont {Giannozzi}}, \bibinfo {author} {\bibfnamefont {O.}~\bibnamefont {Andreussi}}, \bibinfo {author} {\bibfnamefont {T.}~\bibnamefont {Brumme}}, \bibinfo {author} {\bibfnamefont {O.}~\bibnamefont {Bunau}}, \bibinfo {author} {\bibfnamefont {M.}~\bibnamefont {Buongiorno~Nardelli}}, \bibinfo {author} {\bibfnamefont {M.}~\bibnamefont {Calandra}}, \bibinfo {author} {\bibfnamefont {R.}~\bibnamefont {Car}}, \bibinfo {author} {\bibfnamefont {C.}~\bibnamefont {Cavazzoni}}, \bibinfo {author} {\bibfnamefont {D.}~\bibnamefont {Ceresoli}}, \bibinfo {author} {\bibfnamefont {M.}~\bibnamefont {Cococcioni}}, \bibinfo {author} {\bibfnamefont {N.}~\bibnamefont {Colonna}}, \bibinfo {author} {\bibfnamefont {I.}~\bibnamefont {Carnimeo}}, \bibinfo {author} {\bibfnamefont {A.}~\bibnamefont {Dal~Corso}}, \bibinfo {author} {\bibfnamefont {S.}~\bibnamefont {de~Gironcoli}}, \bibinfo {author} {\bibfnamefont {P.}~\bibnamefont {Delugas}}, \bibinfo {author} {\bibfnamefont
  {R.~A.}\ \bibnamefont {DiStasio}}, \bibinfo {author} {\bibfnamefont {A.}~\bibnamefont {Ferretti}}, \bibinfo {author} {\bibfnamefont {A.}~\bibnamefont {Floris}}, \bibinfo {author} {\bibfnamefont {G.}~\bibnamefont {Fratesi}}, \bibinfo {author} {\bibfnamefont {G.}~\bibnamefont {Fugallo}}, \bibinfo {author} {\bibfnamefont {R.}~\bibnamefont {Gebauer}}, \bibinfo {author} {\bibfnamefont {U.}~\bibnamefont {Gerstmann}}, \bibinfo {author} {\bibfnamefont {F.}~\bibnamefont {Giustino}}, \bibinfo {author} {\bibfnamefont {T.}~\bibnamefont {Gorni}}, \bibinfo {author} {\bibfnamefont {J.}~\bibnamefont {Jia}}, \bibinfo {author} {\bibfnamefont {M.}~\bibnamefont {Kawamura}}, \bibinfo {author} {\bibfnamefont {H.-Y.}\ \bibnamefont {Ko}}, \bibinfo {author} {\bibfnamefont {A.}~\bibnamefont {Kokalj}}, \bibinfo {author} {\bibfnamefont {E.}~\bibnamefont {Küçükbenli}}, \bibinfo {author} {\bibfnamefont {M.}~\bibnamefont {Lazzeri}}, \bibinfo {author} {\bibfnamefont {M.}~\bibnamefont {Marsili}}, \bibinfo {author} {\bibfnamefont
  {N.}~\bibnamefont {Marzari}}, \bibinfo {author} {\bibfnamefont {F.}~\bibnamefont {Mauri}}, \bibinfo {author} {\bibfnamefont {N.~L.}\ \bibnamefont {Nguyen}}, \bibinfo {author} {\bibfnamefont {H.-V.}\ \bibnamefont {Nguyen}}, \bibinfo {author} {\bibfnamefont {A.}~\bibnamefont {Otero-de-la Roza}}, \bibinfo {author} {\bibfnamefont {L.}~\bibnamefont {Paulatto}}, \bibinfo {author} {\bibfnamefont {S.}~\bibnamefont {Poncé}}, \bibinfo {author} {\bibfnamefont {D.}~\bibnamefont {Rocca}}, \bibinfo {author} {\bibfnamefont {R.}~\bibnamefont {Sabatini}}, \bibinfo {author} {\bibfnamefont {B.}~\bibnamefont {Santra}}, \bibinfo {author} {\bibfnamefont {M.}~\bibnamefont {Schlipf}}, \bibinfo {author} {\bibfnamefont {A.~P.}\ \bibnamefont {Seitsonen}}, \bibinfo {author} {\bibfnamefont {A.}~\bibnamefont {Smogunov}}, \bibinfo {author} {\bibfnamefont {I.}~\bibnamefont {Timrov}}, \bibinfo {author} {\bibfnamefont {T.}~\bibnamefont {Thonhauser}}, \bibinfo {author} {\bibfnamefont {P.}~\bibnamefont {Umari}}, \bibinfo {author}
  {\bibfnamefont {N.}~\bibnamefont {Vast}}, \bibinfo {author} {\bibfnamefont {X.}~\bibnamefont {Wu}},\ and\ \bibinfo {author} {\bibfnamefont {S.}~\bibnamefont {Baroni}},\ }\bibfield  {title} {\bibinfo {title} {Advanced capabilities for materials modelling with {Q}uantum {ESPRESSO}},\ }\href {https://doi.org/10.1088/1361-648X/aa8f79} {\bibfield  {journal} {\bibinfo  {journal} {Journal of Physics: Condensed Matter}\ }\textbf {\bibinfo {volume} {29}},\ \bibinfo {pages} {465901} (\bibinfo {year} {2017})}\BibitemShut {NoStop}%
\bibitem [{\citenamefont {Iba\~nez{-}Azpiroz}\ \emph {et~al.}(2018)\citenamefont {Iba\~nez{-}Azpiroz}, \citenamefont {Tsirkin},\ and\ \citenamefont {Souza}}]{PhysRevB.97.245143}%
  \BibitemOpen
  \bibfield  {author} {\bibinfo {author} {\bibfnamefont {J.}~\bibnamefont {Iba\~nez{-}Azpiroz}}, \bibinfo {author} {\bibfnamefont {S.~S.}\ \bibnamefont {Tsirkin}},\ and\ \bibinfo {author} {\bibfnamefont {I.}~\bibnamefont {Souza}},\ }\bibfield  {title} {\bibinfo {title} {Ab initio calculation of the shift photocurrent by {W}annier interpolation},\ }\href {https://doi.org/10.1103/PhysRevB.97.245143} {\bibfield  {journal} {\bibinfo  {journal} {Phys. Rev. B}\ }\textbf {\bibinfo {volume} {97}},\ \bibinfo {pages} {245143} (\bibinfo {year} {2018})}\BibitemShut {NoStop}%
\bibitem [{\citenamefont {Puente-Uriona}\ \emph {et~al.}(2023)\citenamefont {Puente-Uriona}, \citenamefont {Tsirkin}, \citenamefont {Souza},\ and\ \citenamefont {Iba\~nez{-}Azpiroz}}]{PhysRevB.107.205204}%
  \BibitemOpen
  \bibfield  {author} {\bibinfo {author} {\bibfnamefont {A.~R.}\ \bibnamefont {Puente-Uriona}}, \bibinfo {author} {\bibfnamefont {S.~S.}\ \bibnamefont {Tsirkin}}, \bibinfo {author} {\bibfnamefont {I.}~\bibnamefont {Souza}},\ and\ \bibinfo {author} {\bibfnamefont {J.}~\bibnamefont {Iba\~nez{-}Azpiroz}},\ }\bibfield  {title} {\bibinfo {title} {Ab initio study of the nonlinear optical properties and dc photocurrent of the weyl semimetal tairte\textsubscript{4}},\ }\href {https://doi.org/10.1103/PhysRevB.107.205204} {\bibfield  {journal} {\bibinfo  {journal} {Phys. Rev. B}\ }\textbf {\bibinfo {volume} {107}},\ \bibinfo {pages} {205204} (\bibinfo {year} {2023})}\BibitemShut {NoStop}%
\bibitem [{\citenamefont {Iba\~nez{-}Azpiroz}\ \emph {et~al.}(2022)\citenamefont {Iba\~nez{-}Azpiroz}, \citenamefont {de~Juan},\ and\ \citenamefont {Souza}}]{ibanez2022assessing}%
  \BibitemOpen
  \bibfield  {author} {\bibinfo {author} {\bibfnamefont {J.}~\bibnamefont {Iba\~nez{-}Azpiroz}}, \bibinfo {author} {\bibfnamefont {F.}~\bibnamefont {de~Juan}},\ and\ \bibinfo {author} {\bibfnamefont {I.}~\bibnamefont {Souza}},\ }\bibfield  {title} {\bibinfo {title} {Assessing the role of interatomic position matrix elements in tight-binding calculations of optical properties},\ }\href {https://www.scipost.org/10.21468/SciPostPhys.12.2.070?acad_field_slug=chemistry} {\bibfield  {journal} {\bibinfo  {journal} {SciPost Physics}\ }\textbf {\bibinfo {volume} {12}},\ \bibinfo {pages} {070} (\bibinfo {year} {2022})}\BibitemShut {NoStop}%
\bibitem [{\citenamefont {Sakamoto}\ \emph {et~al.}(2009)\citenamefont {Sakamoto}, \citenamefont {Oda}, \citenamefont {Kimura}, \citenamefont {Miyamoto}, \citenamefont {Tsujikawa}, \citenamefont {Imai}, \citenamefont {Ueno}, \citenamefont {Namatame}, \citenamefont {Taniguchi}, \citenamefont {Eriksson},\ and\ \citenamefont {Uhrberg}}]{PhysRevLett.102.096805}%
  \BibitemOpen
  \bibfield  {author} {\bibinfo {author} {\bibfnamefont {K.}~\bibnamefont {Sakamoto}}, \bibinfo {author} {\bibfnamefont {T.}~\bibnamefont {Oda}}, \bibinfo {author} {\bibfnamefont {A.}~\bibnamefont {Kimura}}, \bibinfo {author} {\bibfnamefont {K.}~\bibnamefont {Miyamoto}}, \bibinfo {author} {\bibfnamefont {M.}~\bibnamefont {Tsujikawa}}, \bibinfo {author} {\bibfnamefont {A.}~\bibnamefont {Imai}}, \bibinfo {author} {\bibfnamefont {N.}~\bibnamefont {Ueno}}, \bibinfo {author} {\bibfnamefont {H.}~\bibnamefont {Namatame}}, \bibinfo {author} {\bibfnamefont {M.}~\bibnamefont {Taniguchi}}, \bibinfo {author} {\bibfnamefont {P.~E.~J.}\ \bibnamefont {Eriksson}},\ and\ \bibinfo {author} {\bibfnamefont {R.~I.~G.}\ \bibnamefont {Uhrberg}},\ }\bibfield  {title} {\bibinfo {title} {Abrupt rotation of the {R}ashba spin to the direction perpendicular to the surface},\ }\href {https://doi.org/10.1103/PhysRevLett.102.096805} {\bibfield  {journal} {\bibinfo  {journal} {Phys. Rev. Lett.}\ }\textbf {\bibinfo {volume} {102}},\ \bibinfo
  {pages} {096805} (\bibinfo {year} {2009})}\BibitemShut {NoStop}%
\bibitem [{\citenamefont {Stolwijk}\ \emph {et~al.}(2017)\citenamefont {Stolwijk}, \citenamefont {Schmidt}, \citenamefont {Sakamoto}, \citenamefont {Kr\"uger},\ and\ \citenamefont {Donath}}]{PhysRevMaterials.1.064604}%
  \BibitemOpen
  \bibfield  {author} {\bibinfo {author} {\bibfnamefont {S.~D.}\ \bibnamefont {Stolwijk}}, \bibinfo {author} {\bibfnamefont {A.~B.}\ \bibnamefont {Schmidt}}, \bibinfo {author} {\bibfnamefont {K.}~\bibnamefont {Sakamoto}}, \bibinfo {author} {\bibfnamefont {P.}~\bibnamefont {Kr\"uger}},\ and\ \bibinfo {author} {\bibfnamefont {M.}~\bibnamefont {Donath}},\ }\bibfield  {title} {\bibinfo {title} {Valley spin polarization of {Tl/Si(111)}},\ }\href {https://doi.org/10.1103/PhysRevMaterials.1.064604} {\bibfield  {journal} {\bibinfo  {journal} {Phys. Rev. Mater.}\ }\textbf {\bibinfo {volume} {1}},\ \bibinfo {pages} {064604} (\bibinfo {year} {2017})}\BibitemShut {NoStop}%
\bibitem [{\citenamefont {Liu}\ and\ \citenamefont {Chang}(2009)}]{PhysRevB.80.241304}%
  \BibitemOpen
  \bibfield  {author} {\bibinfo {author} {\bibfnamefont {M.-H.}\ \bibnamefont {Liu}}\ and\ \bibinfo {author} {\bibfnamefont {C.-R.}\ \bibnamefont {Chang}},\ }\bibfield  {title} {\bibinfo {title} {Upstanding {R}ashba spin in honeycomb lattices: Electrically reversible surface spin polarization},\ }\href {https://doi.org/10.1103/PhysRevB.80.241304} {\bibfield  {journal} {\bibinfo  {journal} {Phys. Rev. B}\ }\textbf {\bibinfo {volume} {80}},\ \bibinfo {pages} {241304} (\bibinfo {year} {2009})}\BibitemShut {NoStop}%
\bibitem [{\citenamefont {Iba\~nez{-}Azpiroz}\ \emph {et~al.}(2011)\citenamefont {Iba\~nez{-}Azpiroz}, \citenamefont {Eiguren},\ and\ \citenamefont {Bergara}}]{PhysRevB.84.125435}%
  \BibitemOpen
  \bibfield  {author} {\bibinfo {author} {\bibfnamefont {J.}~\bibnamefont {Iba\~nez{-}Azpiroz}}, \bibinfo {author} {\bibfnamefont {A.}~\bibnamefont {Eiguren}},\ and\ \bibinfo {author} {\bibfnamefont {A.}~\bibnamefont {Bergara}},\ }\bibfield  {title} {\bibinfo {title} {Relativistic effects and fully spin-polarized {F}ermi surface at the {Tl/Si(111)} surface},\ }\href {https://doi.org/10.1103/PhysRevB.84.125435} {\bibfield  {journal} {\bibinfo  {journal} {Phys. Rev. B}\ }\textbf {\bibinfo {volume} {84}},\ \bibinfo {pages} {125435} (\bibinfo {year} {2011})}\BibitemShut {NoStop}%
\bibitem [{\citenamefont {Glass}\ \emph {et~al.}(1974)\citenamefont {Glass}, \citenamefont {von~der Linde},\ and\ \citenamefont {Negran}}]{glass-apl74}%
  \BibitemOpen
  \bibfield  {author} {\bibinfo {author} {\bibfnamefont {A.~M.}\ \bibnamefont {Glass}}, \bibinfo {author} {\bibfnamefont {D.}~\bibnamefont {von~der Linde}},\ and\ \bibinfo {author} {\bibfnamefont {T.~J.}\ \bibnamefont {Negran}},\ }\bibfield  {title} {\bibinfo {title} {High-voltage bulk photovoltaic effect and the photorefractive process in {LiNbO}$_3$},\ }\href {https://doi.org/10.1063/1.1655453} {\bibfield  {journal} {\bibinfo  {journal} {Appl. Phys. Lett.}\ }\textbf {\bibinfo {volume} {25}},\ \bibinfo {pages} {233} (\bibinfo {year} {1974})}\BibitemShut {NoStop}%
\bibitem [{\citenamefont {Tan}\ \emph {et~al.}(2016)\citenamefont {Tan}, \citenamefont {Zheng}, \citenamefont {Young}, \citenamefont {Wang}, \citenamefont {Liu},\ and\ \citenamefont {Rappe}}]{tan-cm16}%
  \BibitemOpen
  \bibfield  {author} {\bibinfo {author} {\bibfnamefont {L.~Z.}\ \bibnamefont {Tan}}, \bibinfo {author} {\bibfnamefont {F.}~\bibnamefont {Zheng}}, \bibinfo {author} {\bibfnamefont {S.~M.}\ \bibnamefont {Young}}, \bibinfo {author} {\bibfnamefont {F.}~\bibnamefont {Wang}}, \bibinfo {author} {\bibfnamefont {S.}~\bibnamefont {Liu}},\ and\ \bibinfo {author} {\bibfnamefont {A.~M.}\ \bibnamefont {Rappe}},\ }\bibfield  {title} {\bibinfo {title} {Shift current bulk photovoltaic effect in polar materials -- hybrid and oxide perovskites and beyond},\ }\href {https://doi.org/10.1038/npjcompumats.2016.26} {\bibfield  {journal} {\bibinfo  {journal} {npj Comput. Mater.}\ }\textbf {\bibinfo {volume} {2}},\ \bibinfo {pages} {16026} (\bibinfo {year} {2016})}\BibitemShut {NoStop}%
\bibitem [{\citenamefont {W.~T. H.~Koch}\ and\ \citenamefont {Würfel}(1976)}]{measurement3}%
  \BibitemOpen
  \bibfield  {author} {\bibinfo {author} {\bibfnamefont {W.~R.}\ \bibnamefont {W.~T. H.~Koch}, \bibfnamefont {R.~Munser}}\ and\ \bibinfo {author} {\bibfnamefont {P.}~\bibnamefont {Würfel}},\ }\bibfield  {title} {\bibinfo {title} {Anomalous photovoltage in {BaTiO}\textsubscript{3}},\ }\href {https://doi.org/10.1080/00150197608236596} {\bibfield  {journal} {\bibinfo  {journal} {Ferroelectrics}\ }\textbf {\bibinfo {volume} {13}},\ \bibinfo {pages} {305} (\bibinfo {year} {1976})}\BibitemShut {NoStop}%
\bibitem [{\citenamefont {Xu}\ \emph {et~al.}(2021{\natexlab{b}})\citenamefont {Xu}, \citenamefont {Wang}, \citenamefont {Zhou},\ and\ \citenamefont {Li}}]{Xu2021}%
  \BibitemOpen
  \bibfield  {author} {\bibinfo {author} {\bibfnamefont {H.}~\bibnamefont {Xu}}, \bibinfo {author} {\bibfnamefont {H.}~\bibnamefont {Wang}}, \bibinfo {author} {\bibfnamefont {J.}~\bibnamefont {Zhou}},\ and\ \bibinfo {author} {\bibfnamefont {J.}~\bibnamefont {Li}},\ }\bibfield  {title} {\bibinfo {title} {Pure spin photocurrent in non-centrosymmetric crystals: bulk spin photovoltaic effect},\ }\href {https://doi.org/10.1038/s41467-021-24541-7} {\bibfield  {journal} {\bibinfo  {journal} {Nature Communications}\ }\textbf {\bibinfo {volume} {12}},\ \bibinfo {pages} {4330} (\bibinfo {year} {2021}{\natexlab{b}})}\BibitemShut {NoStop}%
\bibitem [{\citenamefont {Bihlmayer}\ \emph {et~al.}(2018)\citenamefont {Bihlmayer}, \citenamefont {Buhl}, \citenamefont {Dupe}, \citenamefont {Fernandes}, \citenamefont {Freimuth}, \citenamefont {Gayles}, \citenamefont {Heinze}, \citenamefont {Kiselev}, \citenamefont {Lounis}, \citenamefont {Mokrousov},\ and\ \citenamefont {Bl{\"u}gel}}]{bihlmayer_magnetic_nodate}%
  \BibitemOpen
  \bibfield  {author} {\bibinfo {author} {\bibfnamefont {G.}~\bibnamefont {Bihlmayer}}, \bibinfo {author} {\bibfnamefont {P.~M.}\ \bibnamefont {Buhl}}, \bibinfo {author} {\bibfnamefont {B.}~\bibnamefont {Dupe}}, \bibinfo {author} {\bibfnamefont {I.~L.}\ \bibnamefont {Fernandes}}, \bibinfo {author} {\bibfnamefont {F.}~\bibnamefont {Freimuth}}, \bibinfo {author} {\bibfnamefont {J.}~\bibnamefont {Gayles}}, \bibinfo {author} {\bibfnamefont {S.}~\bibnamefont {Heinze}}, \bibinfo {author} {\bibfnamefont {N.}~\bibnamefont {Kiselev}}, \bibinfo {author} {\bibfnamefont {S.}~\bibnamefont {Lounis}}, \bibinfo {author} {\bibfnamefont {Y.}~\bibnamefont {Mokrousov}},\ and\ \bibinfo {author} {\bibfnamefont {S.}~\bibnamefont {Bl{\"u}gel}},\ }\href@noop {} {\bibfield  {journal} {\bibinfo  {journal} {Psi{-}k Newsletter {-} Highlights}\ }\textbf {\bibinfo {volume} {139}},\ \bibinfo {pages} {40} (\bibinfo {year} {2018})}\BibitemShut {NoStop}%
\bibitem [{\citenamefont {Bihlmayer}\ \emph {et~al.}(2007)\citenamefont {Bihlmayer}, \citenamefont {Bl\"ugel},\ and\ \citenamefont {Chulkov}}]{PhysRevB.75.195414}%
  \BibitemOpen
  \bibfield  {author} {\bibinfo {author} {\bibfnamefont {G.}~\bibnamefont {Bihlmayer}}, \bibinfo {author} {\bibfnamefont {S.}~\bibnamefont {Bl\"ugel}},\ and\ \bibinfo {author} {\bibfnamefont {E.~V.}\ \bibnamefont {Chulkov}},\ }\bibfield  {title} {\bibinfo {title} {Enhanced rashba spin-orbit splitting in {$\mathrm{Bi}/\mathrm{Ag}(111)$ and $\mathrm{Pb}/\mathrm{Ag}(111)$} surface alloys from first principles},\ }\href {https://doi.org/10.1103/PhysRevB.75.195414} {\bibfield  {journal} {\bibinfo  {journal} {Phys. Rev. B}\ }\textbf {\bibinfo {volume} {75}},\ \bibinfo {pages} {195414} (\bibinfo {year} {2007})}\BibitemShut {NoStop}%
\bibitem [{\citenamefont {Kato}\ \emph {et~al.}(2004)\citenamefont {Kato}, \citenamefont {Myers}, \citenamefont {Gossard},\ and\ \citenamefont {Awschalom}}]{SPHmeassurement}%
  \BibitemOpen
  \bibfield  {author} {\bibinfo {author} {\bibfnamefont {Y.~K.}\ \bibnamefont {Kato}}, \bibinfo {author} {\bibfnamefont {R.~C.}\ \bibnamefont {Myers}}, \bibinfo {author} {\bibfnamefont {A.~C.}\ \bibnamefont {Gossard}},\ and\ \bibinfo {author} {\bibfnamefont {D.~D.}\ \bibnamefont {Awschalom}},\ }\bibfield  {title} {\bibinfo {title} {Observation of the spin {H}all effect in semiconductors},\ }\href {https://doi.org/10.1126/science.1105514} {\bibfield  {journal} {\bibinfo  {journal} {Science}\ }\textbf {\bibinfo {volume} {306}},\ \bibinfo {pages} {1910} (\bibinfo {year} {2004})},\ \Eprint {https://arxiv.org/abs/https://www.science.org/doi/pdf/10.1126/science.1105514} {https://www.science.org/doi/pdf/10.1126/science.1105514} \BibitemShut {NoStop}%
\bibitem [{\citenamefont {Sinova}\ \emph {et~al.}(2004)\citenamefont {Sinova}, \citenamefont {Culcer}, \citenamefont {Niu}, \citenamefont {Sinitsyn}, \citenamefont {Jungwirth},\ and\ \citenamefont {MacDonald}}]{Sinova2004}%
  \BibitemOpen
  \bibfield  {author} {\bibinfo {author} {\bibfnamefont {J.}~\bibnamefont {Sinova}}, \bibinfo {author} {\bibfnamefont {D.}~\bibnamefont {Culcer}}, \bibinfo {author} {\bibfnamefont {Q.}~\bibnamefont {Niu}}, \bibinfo {author} {\bibfnamefont {N.~A.}\ \bibnamefont {Sinitsyn}}, \bibinfo {author} {\bibfnamefont {T.}~\bibnamefont {Jungwirth}},\ and\ \bibinfo {author} {\bibfnamefont {A.~H.}\ \bibnamefont {MacDonald}},\ }\bibfield  {title} {\bibinfo {title} {Universal intrinsic spin {H}all effect},\ }\href {https://doi.org/10.1103/PhysRevLett.92.126603} {\bibfield  {journal} {\bibinfo  {journal} {Phys. Rev. Lett.}\ }\textbf {\bibinfo {volume} {92}},\ \bibinfo {pages} {126603} (\bibinfo {year} {2004})}\BibitemShut {NoStop}%
\bibitem [{\citenamefont {Schliemann}(2006)}]{Schliemann2006}%
  \BibitemOpen
  \bibfield  {author} {\bibinfo {author} {\bibfnamefont {J.}~\bibnamefont {Schliemann}},\ }\bibfield  {title} {\bibinfo {title} {Spin {H}all effect},\ }\href {https://doi.org/10.1142/S021797920603370X} {\bibfield  {journal} {\bibinfo  {journal} {International Journal of Modern Physics B}\ }\textbf {\bibinfo {volume} {20}},\ \bibinfo {pages} {1015} (\bibinfo {year} {2006})}\BibitemShut {NoStop}%
\bibitem [{\citenamefont {Perdew}\ \emph {et~al.}(1996)\citenamefont {Perdew}, \citenamefont {Burke},\ and\ \citenamefont {Ernzerhof}}]{PhysRevLett.77.3865}%
  \BibitemOpen
  \bibfield  {author} {\bibinfo {author} {\bibfnamefont {J.~P.}\ \bibnamefont {Perdew}}, \bibinfo {author} {\bibfnamefont {K.}~\bibnamefont {Burke}},\ and\ \bibinfo {author} {\bibfnamefont {M.}~\bibnamefont {Ernzerhof}},\ }\bibfield  {title} {\bibinfo {title} {Generalized gradient approximation made simple},\ }\href {https://doi.org/10.1103/PhysRevLett.77.3865} {\bibfield  {journal} {\bibinfo  {journal} {Phys. Rev. Lett.}\ }\textbf {\bibinfo {volume} {77}},\ \bibinfo {pages} {3865} (\bibinfo {year} {1996})}\BibitemShut {NoStop}%
\bibitem [{\citenamefont {Pizzi}\ \emph {et~al.}(2020)\citenamefont {Pizzi}, \citenamefont {Vitale}, \citenamefont {Arita}, \citenamefont {Bl{\"u}gel}, \citenamefont {Freimuth}, \citenamefont {G{\'e}ranton}, \citenamefont {Gibertini}, \citenamefont {Gresch}, \citenamefont {Johnson}, \citenamefont {Koretsune}, \citenamefont {{Iba{\~n}ez-Azpiroz}}, \citenamefont {Lee}, \citenamefont {Lihm}, \citenamefont {Marchand}, \citenamefont {Marrazzo}, \citenamefont {Mokrousov}, \citenamefont {Mustafa}, \citenamefont {Nohara}, \citenamefont {Nomura}, \citenamefont {Paulatto}, \citenamefont {Ponc{\'e}}, \citenamefont {Ponweiser}, \citenamefont {Qiao}, \citenamefont {Th{\"o}le}, \citenamefont {Tsirkin}, \citenamefont {Wierzbowska}, \citenamefont {Marzari}, \citenamefont {Vanderbilt}, \citenamefont {Souza}, \citenamefont {Mostofi},\ and\ \citenamefont {Yates}}]{MOSTOFI20142309}%
  \BibitemOpen
  \bibfield  {author} {\bibinfo {author} {\bibfnamefont {G.}~\bibnamefont {Pizzi}}, \bibinfo {author} {\bibfnamefont {V.}~\bibnamefont {Vitale}}, \bibinfo {author} {\bibfnamefont {R.}~\bibnamefont {Arita}}, \bibinfo {author} {\bibfnamefont {S.}~\bibnamefont {Bl{\"u}gel}}, \bibinfo {author} {\bibfnamefont {F.}~\bibnamefont {Freimuth}}, \bibinfo {author} {\bibfnamefont {G.}~\bibnamefont {G{\'e}ranton}}, \bibinfo {author} {\bibfnamefont {M.}~\bibnamefont {Gibertini}}, \bibinfo {author} {\bibfnamefont {D.}~\bibnamefont {Gresch}}, \bibinfo {author} {\bibfnamefont {C.}~\bibnamefont {Johnson}}, \bibinfo {author} {\bibfnamefont {T.}~\bibnamefont {Koretsune}}, \bibinfo {author} {\bibfnamefont {J.}~\bibnamefont {{Iba{\~n}ez-Azpiroz}}}, \bibinfo {author} {\bibfnamefont {H.}~\bibnamefont {Lee}}, \bibinfo {author} {\bibfnamefont {J.-M.}\ \bibnamefont {Lihm}}, \bibinfo {author} {\bibfnamefont {D.}~\bibnamefont {Marchand}}, \bibinfo {author} {\bibfnamefont {A.}~\bibnamefont {Marrazzo}}, \bibinfo {author} {\bibfnamefont
  {Y.}~\bibnamefont {Mokrousov}}, \bibinfo {author} {\bibfnamefont {J.~I.}\ \bibnamefont {Mustafa}}, \bibinfo {author} {\bibfnamefont {Y.}~\bibnamefont {Nohara}}, \bibinfo {author} {\bibfnamefont {Y.}~\bibnamefont {Nomura}}, \bibinfo {author} {\bibfnamefont {L.}~\bibnamefont {Paulatto}}, \bibinfo {author} {\bibfnamefont {S.}~\bibnamefont {Ponc{\'e}}}, \bibinfo {author} {\bibfnamefont {T.}~\bibnamefont {Ponweiser}}, \bibinfo {author} {\bibfnamefont {J.}~\bibnamefont {Qiao}}, \bibinfo {author} {\bibfnamefont {F.}~\bibnamefont {Th{\"o}le}}, \bibinfo {author} {\bibfnamefont {S.~S.}\ \bibnamefont {Tsirkin}}, \bibinfo {author} {\bibfnamefont {M.}~\bibnamefont {Wierzbowska}}, \bibinfo {author} {\bibfnamefont {N.}~\bibnamefont {Marzari}}, \bibinfo {author} {\bibfnamefont {D.}~\bibnamefont {Vanderbilt}}, \bibinfo {author} {\bibfnamefont {I.}~\bibnamefont {Souza}}, \bibinfo {author} {\bibfnamefont {A.~A.}\ \bibnamefont {Mostofi}},\ and\ \bibinfo {author} {\bibfnamefont {J.~R.}\ \bibnamefont {Yates}},\ }\bibfield
  {title} {\bibinfo {title} {{Wannier90} as a community code: New features and applications},\ }\href {https://doi.org/10.1088/1361-648X/ab51ff} {\bibfield  {journal} {\bibinfo  {journal} {Journal of Physics: Condensed Matter}\ }\textbf {\bibinfo {volume} {32}},\ \bibinfo {pages} {165902} (\bibinfo {year} {2020})}\BibitemShut {NoStop}%
\bibitem [{\citenamefont {Lee}\ and\ \citenamefont {Choi}(2012)}]{PhysRevB.86.045437}%
  \BibitemOpen
  \bibfield  {author} {\bibinfo {author} {\bibfnamefont {H.}~\bibnamefont {Lee}}\ and\ \bibinfo {author} {\bibfnamefont {H.~J.}\ \bibnamefont {Choi}},\ }\bibfield  {title} {\bibinfo {title} {Role of $d$ orbitals in the rashba-type spin splitting for noble-metal surfaces},\ }\href {https://doi.org/10.1103/PhysRevB.86.045437} {\bibfield  {journal} {\bibinfo  {journal} {Phys. Rev. B}\ }\textbf {\bibinfo {volume} {86}},\ \bibinfo {pages} {045437} (\bibinfo {year} {2012})}\BibitemShut {NoStop}%
\bibitem [{\citenamefont {Tsirkin}(2021)}]{Tsirkin2021}%
  \BibitemOpen
  \bibfield  {author} {\bibinfo {author} {\bibfnamefont {S.~S.}\ \bibnamefont {Tsirkin}},\ }\bibfield  {title} {\bibinfo {title} {High performance {W}annier interpolation of berry curvature and related quantities with {WannierBerri} code},\ }\href {https://doi.org/10.1038/s41524-021-00498-5} {\bibfield  {journal} {\bibinfo  {journal} {npj Computational Materials}\ }\textbf {\bibinfo {volume} {7}},\ \bibinfo {pages} {33} (\bibinfo {year} {2021})}\BibitemShut {NoStop}%
\bibitem [{\citenamefont {Fei}\ \emph {et~al.}(2020)\citenamefont {Fei}, \citenamefont {Tan},\ and\ \citenamefont {Rappe}}]{PhysRevB.101.045104}%
  \BibitemOpen
  \bibfield  {author} {\bibinfo {author} {\bibfnamefont {R.}~\bibnamefont {Fei}}, \bibinfo {author} {\bibfnamefont {L.~Z.}\ \bibnamefont {Tan}},\ and\ \bibinfo {author} {\bibfnamefont {A.~M.}\ \bibnamefont {Rappe}},\ }\bibfield  {title} {\bibinfo {title} {Shift-current bulk photovoltaic effect influenced by quasiparticle and exciton},\ }\href {https://doi.org/10.1103/PhysRevB.101.045104} {\bibfield  {journal} {\bibinfo  {journal} {Phys. Rev. B}\ }\textbf {\bibinfo {volume} {101}},\ \bibinfo {pages} {045104} (\bibinfo {year} {2020})}\BibitemShut {NoStop}%
\end{thebibliography}%

\newpage

\setcounter{equation}{0}
\setcounter{figure}{0}
\renewcommand{\thesubsection}{S\arabic{subsection}}
\renewcommand{\thetable}{S\arabic{table}}
\renewcommand{\thefigure}{S\arabic{figure}}
\renewcommand{\theequation}{S\arabic{equation}}
\onecolumngrid

\section{Supplementary material} \label{sec: supplementary}

\subsection{Expressions of charge shift and charge/spin injection photoconductivities}

This sections contains the expressions of the tensors that appear on the main text, all expressed in the length gauge as written in \cite{PhysRevB.105.045201, PhysRevB.61.5337}.

\subsubsection{Shift}

\paragraph{Charge.}
The expression for the shift photoconductivity reads 
\begin{eqnarray}
    \sigma^{abc}(\omega) = &&-\frac{i\pi{e}^3}{2\hbar^2}\int \frac{d\mathbf{k}}{(2\pi)^3}\sum_{mn}f_{nm}\sigma^{abc}(\mathbf{k})\delta\left(\omega_{nm}+\omega\right). \label{sup:sc}
\end{eqnarray}
Here $a,b$ and $c$ are Cartesian components,
$m$ and $n$ are band indexes,
$f_{nm}=f_{n}-f_{m}$ is the occupation factor difference, 
$\omega_{mn}=\omega_{m}-\omega_{n}$ is the energy gap of the bands involved 
and $\sigma^{abc}(\mathbf{k})=r_{mn}^{b;a}r^{c}_{nm} - r^{c;a}_{nm}r^{b}_{mn}$ is the k-resolved tensor 
which contains interband Berry connection $r^{a}_{mn} = i\braket{u_m  | u_n}$ 
and its generalized derivative $r_{mn}^{b;a}=\partial_a r_{mn}^{b}-i\left(A_{mm}^{a}-A_{nn}^{a}\right)r_{mn}^{b}$
with $A_{nn}^{a}=i\bra{n}\partial_{a}\ket{n}$ the intraband Berry connection. 

For $\mathcal{T}$-invariant systems the shift current results from the real part of the symmetric combination $\text{Re}(\sigma^{abc} + \sigma^{acb})$ corresponding to the generation of current under linearly polarized light.

\subsubsection{Injection}

\paragraph{Charge.}
The charge injection tensor can be written as 
\begin{eqnarray}
    \eta^{abc}(\omega) = -\tau\frac{\pi q^3}{\hbar^2 \omega^2 V}\sum_{\mathbf{k},m,n}f_{mn} \eta^{abc}_{mn}(\mathbf{k}) \delta(\omega+\omega_{mn}), \label{sup:ic}
\end{eqnarray}
with $\eta^{s;abc}_{mn}(\mathbf{k})  = \Delta v^{a}_{mn} v^{b}_{mn} v^{c}_{nm}$
and $\Delta v^{a}_{mn}  = v^{a}_{mm} - v^{a}_{nn}$. 

For $\mathcal{T}$-invariant systems the injection current results from the imaginary part of the antisymmetric combination $\text{Im}(\eta^{abc} - \eta^{acb})$ corresponding to the generation of current under circularly polarized light.

\paragraph{Spin.}
In the spin case, the expression becomes
\begin{eqnarray}
    \xi^{s,abc}(\omega) = -\tau \frac{\pi q^3}{\hbar^2 \omega^2 V}\sum_{\mathbf{k},m,n}f_{mn} \xi^{s,abc}_{mn}(\mathbf{k}) \delta(\omega+\omega_{mn}), \label{sup:spic}
\end{eqnarray}
with $\xi^{s;abc}_{mn}(\mathbf{k})  = \Delta j^{s,a}_{mn} v^{b}_{mn} v^{c}_{nm}$,
$\Delta j^{s,a}_{mn}  = j^{s,a}_{mm} - j^{s,a}_{nn}$ and
$j^{s,a}_{mn} = \bra{u_m}\{v^{a}, S^{s}\}\ket{n}$  the spin-current matrix elements as considered in \cite{PhysRevB.105.045201}.

For $\mathcal{T}$-invariant systems, the shift current results from the real part of the symmetric combination $\text{Re}(\xi^{s,abc} + \xi^{s,acb})$ corresponding to the generation of current under linearly polarized light.

\subsection{Analytical expressions in the Rashba model }

In this section, we provide more details regarding the analytical expressions obtained for the Rashba mode as well as the complete set of charge shift and injection tensor components. The starting point is the linear Rashba Hamiltonian in polar coordinates
\begin{equation}
    H_0 = \left[\begin{matrix}\frac{\rho^{2}}{2 m} & - i \alpha \rho e^{- i \phi}\\i \alpha \rho e^{i \phi} & \frac{\rho^{2}}{2 m}\end{matrix}\right]
\end{equation}
together with an extension comprising of higher order terms that comply with the C$_{3v}$ point group 
\begin{equation}
    V = \left[\begin{matrix}\beta \left(- 3 \rho^{3} \sin^{2}{\left(\phi \right)} \cos{\left(\phi \right)} + \rho^{3} \cos^{3}{\left(\phi \right)}\right) & - i \gamma \rho^{3} e^{- i \phi}\\i \gamma \rho^{3} e^{i \phi} & - \beta \left(- 3 \rho^{3} \sin^{2}{\left(\phi \right)} \cos{\left(\phi \right)} + \rho^{3} \cos^{3}{\left(\phi \right)}\right)\end{matrix}\right]
\end{equation}
which we treat as a perturbation. Up to first order in the perturbation, the eigenstates of $H=H_0 +V$ are
\begin{align}
\ket{+} &= \left[\begin{matrix}- \frac{\sqrt{2} i \left(2 \alpha + \beta \rho^{2} \cos{\left(3 \phi \right)}\right) e^{- i \phi}}{4 \alpha}\\\frac{\sqrt{2} \cdot \left(2 \alpha - \beta \rho^{2} \cos{\left(3 \phi \right)}\right)}{4 \alpha}\end{matrix}\right]
\\
\ket{-} &= \left[\begin{matrix}\frac{\sqrt{2} i \left(2 \alpha - \beta \rho^{2} \cos{\left(3 \phi \right)}\right) e^{- i \phi}}{4 \alpha}\\\frac{\sqrt{2} \cdot \left(2 \alpha + \beta \rho^{2} \cos{\left(3 \phi \right)}\right)}{4 \alpha}\end{matrix}\right],
\end{align}
and the energy difference between the bands is
\begin{equation}
    \omega_{-+} = 2 \rho \left(\alpha + \gamma \rho^{2}\right).
\end{equation}
With that we solve \ref{sup:sc} and \ref{sup:ic}, resulting in
\begin{align}
    \sigma^{xxx}(\mathbf{k}) &= \frac{3 \beta \sin{\left(2 \phi \right)}}{2 \alpha \rho} \\
    \sigma^{xxy}(\mathbf{k}) &= \frac{\beta \left(16 \sin^{6}{\left(\phi \right)} + 12 \sin^{4}{\left(\phi \right)} - 35 \sin^{2}{\left(\phi \right)} - \sin{\left(\phi \right)} \sin{\left(5 \phi \right)} - 16\right)}{16 \alpha \rho} \\
    \sigma^{xyy}(\mathbf{k}) & = \frac{\beta \left(- 16 \sin^{5}{\left(\phi \right)} - 12 \sin^{3}{\left(\phi \right)} + 27 \sin{\left(\phi \right)} + \sin{\left(5 \phi \right)}\right) \cos{\left(\phi \right)}}{8 \alpha \rho} \\
    \sigma^{yxx}(\mathbf{k}) & = \frac{\beta \left(16 \sin^{6}{\left(\phi \right)} - 52 \sin^{4}{\left(\phi \right)} + 21 \sin^{2}{\left(\phi \right)} - \sin{\left(\phi \right)} \sin{\left(5 \phi \right)} - 8\right)}{8 \alpha \rho} \\
    \sigma^{yxy}(\mathbf{k}) &= \frac{\beta \left(- 16 \sin^{5}{\left(\phi \right)} + 52 \sin^{3}{\left(\phi \right)} - 13 \sin{\left(\phi \right)} + \sin{\left(5 \phi \right)}\right) \cos{\left(\phi \right)}}{16 \alpha \rho} \\
     \sigma^{yyy}(\mathbf{k}) &= \frac{3 \beta \cos^{2}{\left(\phi \right)}}{\alpha \rho}
\end{align}
and
\begin{align}
    \eta^{xxy}(\mathbf{k}) &= \frac{- 2  \beta \cos{\left(\phi \right)} \cos{\left(3 \phi \right)}}{4  \rho^{2} \left(\alpha^{2} + 4 \alpha \gamma \rho^{2}\right)} \\
 \eta^{yxy}(\mathbf{k}) &= \frac{- 2  \beta \sin{\left(\phi \right)} \cos{\left(3 \phi \right)}}{4  \rho^{2} \left(\alpha^{2} + 4 \alpha \gamma \rho^{2} \right)}
\end{align}
Note that the shift(injection) photoconductivity is symmetric(antisymmetric) in ($b\leftrightarrow c$), implying $\sigma^{xyx}(\mathbf{k}) = \sigma^{xxy}(\mathbf{k})$,  $\sigma^{yyx}(\mathbf{k}) = \sigma^{yxy}(\mathbf{k})$, $\eta^{xyx}(\mathbf{k}) = -\eta^{xxy}(\mathbf{k})$ and $\eta^{yyx}(\mathbf{k}) = -\eta^{yxy}(\mathbf{k})$.

\subsection{Computational details}

\paragraph{Au(111).} 
%
Density functional theory (DFT) calculations were performed using the Quantum ESPRESSO package \cite{Giannozzi_2009, Giannozzi_2017} with the Perdew-Burke-Ernzerhof (PBE) pseudopotential \cite{PhysRevLett.77.3865} as approximation for the exchange-correlation term. The slab consisted of $7$ unit cells, with an energy cutoff of $45$ Ry and a Monkhorst-Pack grid of $8 \times 8 \times 1$ \textbf{k}-points.
Maximally localized Wannier functions were then constructed from the Bloch eigenstates using the Wannier90 software package \cite{MOSTOFI20142309}, where the surface states were projected onto p$_z$ orbitals \cite{PhysRevB.86.045437}.
Finally, the \textbf{k}-resolved shift and injection photoconductivities were computed following the methodology described in Ref. \cite{PhysRevB.97.245143}.

As the maximally localization process does not preserve crystal symmetries,  the \textbf{k}-resolved quantities were symmetrized in the following way. First, we choose an irreducible Brillouin zone. Then, the rest of the Brillouin zone is generated as ~$\sigma^{abc}(\mathbf{k})=\sum_{\alpha \beta \gamma}M_{a\alpha}M_{b\beta}M_{c\gamma}\sigma^{\alpha\beta\gamma}(M^{-1}\mathbf{k})$ where $M$ is the matrix representation of a crystal symmetry. Finally, the results from all possible choices are averaged. 

\paragraph{Tl/Si(111).}
For the calculations on Tl/Si(111) we proceed in the same manner as for the Au(111) surface. In this case, the simulation parameters are a slab size of 4 unit cells, an energy cutoff of $200$ Ry, a $24 \times 24 \times 1$ \textbf{k}-point grid, and projections with $l = 1, m_r = 1$ for the unoccupied states.
For the calculation of the spectra, the WannierBerri code was employed \cite{Tsirkin2021}.

\subsection{Details on the calculation of the current}
In order to calculate the photocurrent from quadratic optical responses, the expression 
for a sample of thickness $d$ illuminated by  a laser with 
spot size of width $w$
reads~\cite{PhysRevB.101.045104, PhysRevB.108.165418}
\begin{eqnarray}
    J^{a} & = G^{abb}(\omega)\cdot(1-R(\omega))(1- e^{-\alpha^{bb}\cdot d})\cdot w \cdot I^{b},\label{sup:current}
\end{eqnarray}
Above, $I^{b}=c\epsilon_{0} (E^{b})^{2}/2$ is the incident light intensity
for an applied electric field $E^b$,   $R(\omega)$  the reflectivity and
$ \alpha^{bb}(\omega) = \sqrt{2}\frac{\omega}{c}\sqrt{|\epsilon^{bb}|-\text{Re}(\epsilon^{bb})}$.
Finally, 
$
    G^{abb}(\omega)=2\sigma^{abb}(\omega)/c\epsilon_0 \sqrt{\epsilon_r}\cdot\alpha^{bb}(\omega) \label{eq:Glass}
$
stands for the Glass coefficient, which 
quantifies the generation of photocurrents in bulk materials
taking absorption into account~\cite{glass-apl74,tan-cm16}.
In our calculations, we consider a power density $P/A=5\cdot 10^{6}$ W/m$^{2}$ in line with \cite{Ma2019}. 
Additionally, for simplicity we have set $R(\omega)$ = 0, so the magnitude of the calculated current is expected to be overestimated. 

\subsection{Spin injection photoconductivity spectra in Au(111)}
\begin{figure}[h!]
    \centering
    \includegraphics[width=0.5\linewidth]{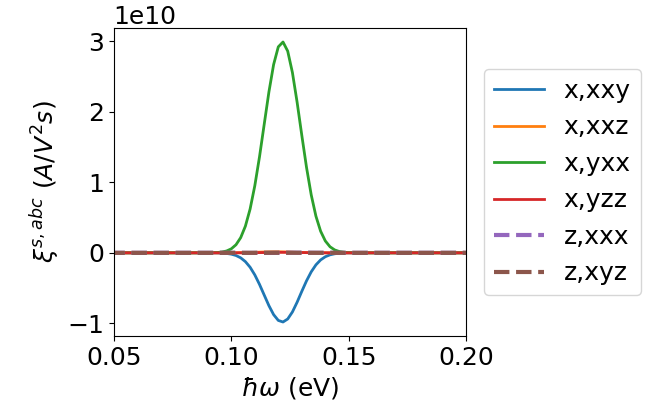}
    \caption{Spin injection photoconductivity spectra in Au(111).}
    \label{fig:spic_au}
\end{figure}

Fig. \ref{fig:spic_au} shows the independent spin injection photoconductivity tensor components for Au(111), where it can be observed that spin-polarized in-plane components dominate over out-of-plane components. As argued in text, this is due to the in-plane spin texture which characterizes Rashba surface states.

\subsection{Symmetric combinations of k-resolved spin tensor}

Fig. 4 in the main text shows symmetric combinations of the k-resolved spin injection photoconductivities
\begin{equation}
    \Xi(\mathbf{k}) = \sum_{s,abc}C^{s,abc}\xi^{s,abc}(\mathbf{k}),
\end{equation}
in the sense that $\Xi(\mathbf{k}) = \Xi(M^{-1}\mathbf{k})$ with $M$ a crystal symmetry. 
The reasoning behind this choice is to show how the spin-polarized injection current along the out-of-plane direction ($z;abc$ components) is correlated to the spin texture which is mostly polarized along z around $K$ in a way that it does not depend on the choice of $K$. 
As the effect of a crystal symmetry on the spin injection photoconductivity is 
\begin{equation}
    \xi^{s,abc}(\mathbf{k})=\sum_{s'a' b'c'}M_{ss'}M_{aa'}M_{bb'}M_{cc'}\xi^{s',a'b'c'}(M^{-1}\mathbf{k}) ,
\end{equation}
then get the relations
\begin{align}
    \Xi(\mathbf{k}) &= \sum_{s,abc}C^{s,abc}\sum_{s'a' b'c'}M_{ss'}M_{aa'}M_{bb'}M_{cc'}\xi^{s',a'b'c'}(M^{-1}\mathbf{k}) \\
    \Xi(\mathbf{k}) &= \Xi(M^{-1}\mathbf{k}) = \sum_{s,abc}C^{s,abc}\xi^{s,abc}(M^{-1}\mathbf{k}) 
\end{align}
form a system of equations which gives as a result the coefficients $C^{s,abc}$, which need not be unique.

In the case of the 3$m$ point group, we find the following combinations to be invariant under $C_{3}$ rotation and $M_{x}$ mirror symmetry operations, one combining tensors that are  out-of-plane and one in-plane spin-polarized;
\begin{align}
    \Xi^{out} (\mathbf{k})\propto & -\xi^{z,xxx}(\mathbf{k}) + \xi^{z,xyy}(\mathbf{k}) + 2\xi^{z,yxy}(\mathbf{k})\\
    \Xi^{in} (\mathbf{k})\propto & -\xi^{x,yxx}(\mathbf{k}) + \xi^{y,xyy}(\mathbf{k}) - \xi^{x,yyy} (\mathbf{k})+ \xi^{y,xxx}(\mathbf{k}).
\end{align}
These combinations correspond to the quantities represented in Fig 4 b).

\end{document}